\documentclass[aps,prb,superscriptaddress,amsmath,amssymb,floatfix,twocolumn,showpacs,amsfonts,longbibliography,reprint]{revtex4-2}
\usepackage{dcolumn}
\usepackage{graphicx}
\usepackage{epstopdf}
\usepackage[caption=false]{subfig} 
\usepackage{epstopdf}
\usepackage{amsmath}
\usepackage{braket}
\usepackage{CJKutf8}
\usepackage{verbatim}
\usepackage{amssymb}
\usepackage{psfrag}
\usepackage{bm}
\usepackage{wrapfig}
\usepackage{makeidx}
\usepackage{bm}
\usepackage{epsf}
\usepackage{multirow}
\usepackage{tikz}
\usepackage{soul}
\usepackage{float}
\usepackage[normalem]{ulem}

\usepackage[colorlinks,linkcolor=blue,urlcolor=blue,citecolor=blue]{hyperref}

\def\pp{\mathbf{p}}
\def\kk{\mathbf{k}}
\def\qq{\mathbf{q}}
\def\bs{\mathbf{S}}
\def\bg{\mathbf{G}}
\def\bq{\mathbf{Q}}

\def\br{\mathbf{r}}

\def\bs{\bm{S}}

\begin{document}
\begin{CJK*}{UTF8}{gbsn}

\title{Multimagnon and multispinon $L_3$-edge RIXS spectra of an effective $\tilde{J}_1-\tilde{J}_2-\tilde{J}_3$ square lattice Heisenberg model}
\author{Kai-Yuan Qi (祁开元)}

\affiliation{State Key Laboratory of Optoelectronic Materials and Technologies, Guangdong Provincial Key Laboratory of Magnetoelectric Physics and Devices, Center for Neutron Science and Technology, School of Physics, Sun Yat-Sen University, Guangzhou 510275, China}
\author{Shangjian Jin}
\affiliation{State Key Laboratory of Optoelectronic Materials and Technologies, Guangdong Provincial Key Laboratory of Magnetoelectric Physics and Devices, Center for Neutron Science and Technology, School of Physics, Sun Yat-Sen University, Guangzhou 510275, China}
 \author{Trinanjan Datta}
 \email[Corresponding author:]{tdatta@augusta.edu}
 \affiliation{Department of Physics and Biophysics, Augusta University, 1120 15th Street, Augusta, Georgia 30912, USA}

\author{Dao-Xin Yao (姚道新)}
\email[Corresponding author:]{yaodaox@mail.sysu.edu.cn}
\affiliation{State Key Laboratory of Optoelectronic Materials and Technologies, Guangdong Provincial Key Laboratory of Magnetoelectric Physics and Devices, Center for Neutron Science and Technology, School of Physics, Sun Yat-Sen University, Guangzhou 510275, China}

\date{\today}	
\begin{abstract}
We investigate the multimagnon and the multispinon $L_3$-edge resonant inelastic x-ray scattering (RIXS) spectra of a spin-1/2 effective $\tilde{J}_1-\tilde{J}_2-\tilde{J}_3$ square lattice Heisenberg model in its N\'{e}el ordered phase. Motivated by the observation of satellite intensity peaks above the single magnon dispersion in the $L$-edge RIXS spectrum, we propose a resonating valence bond (RVB) inspired RIXS mechanism that incorporates the local site ultrashort core-hole lifetime (UCL) expansion. We compute the multimagnon and the multispinon excitations using $\mathcal{O}(1/S)$ interacting spin wave theory and Schwinger boson mean-field theory (SBMFT) formalism, respectively. We treat the x-ray scattering process up to second order in the UCL expansion. Our calculations of two-magnon, bimagnon, and three-magnon RIXS intensities reveal that interacting spin wave theory fails to fully capture all the quantum correlations in the antiferromagnetic ordered phase. However utilizing the SBMFT framework, with a ground state that incorporates features of both the N\'{e}el order and fluctuating RVB components, we demonstrate that a RIXS bond-flipping mechanism provides an alternative deeper physical explanation of the satellite intensities. Specifically, we find that the spin correlation spectra predicted by the fluctuating RVB mechanism aligns with higher order UCL expansion results. We further show that the satellite intensity above the single-magnon mode can originate both from a one-to-three-magnon hybridization vertex process and from condensed spinons exhibiting Higgs mechanism. These features reflect the interplay of quantum fluctuation, entanglement, and gauge interaction effects of quantum magnetism probed by RIXS.
\end{abstract}
\maketitle	
\end{CJK*}

\section{Introduction\label{sec:intro}} 
The importance of studying magnetic excitation properties of cuprates, iron pnictides, and nickelate compounds is now widely acknowledged~\cite{RevModPhys.84.1383,RevModPhys.87.855,sun2023signatures,RevModPhys.87.457,RevModPhys.75.913,PhysRevLett.86.5377}. The underlying magnetic phase acts as a precursor to the development of superconductivity, which is induced upon doping the parent magnetic material. Among the various possible magnetic ordering patterns, the antiferromagnet (AF) phase often arises in the neighborhood of the superconducting phase. Undoped cuprates, typically modeled using the Heisenberg Hamiltonian, are considered to be charge-transfer insulators. In these correlated electronic systems, strong Coulomb repulsion between the electrons lead to a suppression of conductivity which is accompanied with an underlying AF order and its corresponding magnon excitation spectrum. 

Presently, there are numerous theoretical proposals on high-$T_c$ superconductors~\cite{RevModPhys.87.457,RevModPhys.75.913,PhysRevLett.87.047003,lake2002antiferromagnetic}. Relevant to our context of a magnetic material, spin fluctuation in cuprates has been proposed as a possible glue giving rise to Cooper pairing~\cite{RevModPhys.84.1383}. Thus, understanding and characterizing the nature of magnetic excitations is crucial to unraveling the origins of superconductivity in cuprates. A $J_1-J_2-J_3-J_c$ model, where $J_i~(i = 1,2,3)$ represents exchange interactions up to third neighbor and $J_c$ is the cyclic exchange, has been proposed to compute the magnetic properties of an insulating AF Heisenberg square lattice
~\cite{PhysRevLett.86.5377,PhysRevB.66.100403,PhysRevB.79.235130,PhysRevResearch.7.L012053}. The four-spin cyclic exchange interaction $J_c$, arising from a $t/U$ expansion to the fourth-order of the half-filled single-band Hubbard model~\cite{PhysRevB.79.235130}, has been utilized to explain the spectroscopic features observed in both neutron scattering and Raman spectroscopy~\cite{PhysRevB.39.2299}. This model with extended frustrated magnetic interactions has been successfully applied to calculate the magnetic dispersion of cuprates~\cite{PhysRevB.98.125118}. 

There are several spectroscopic techniques available to measure magnetic properties. Inelastic neutron scattering (INS) typically measures single spin-flip and double spin-flip in its transverse and longitudinal channels, respectively~\cite{PhysRevLett.86.5377,dalla2015fractional}. Raman scattering is able to detect multimagnon excitations limited to $\qq \sim 0$~\cite{RevModPhys.79.175}. Anisotropic spin interactions and multipolar coupling in magnetic materials can be detected via nuclear magnetic resonance (NMR)~\cite{PhysRevLett.75.2212}. While the aforementioned experimental techniques have their merits, resonant inelastic x-ray scattering (RIXS) spectroscopy, baring resolution issues, can in principle access much wider ranges of energy and momentum. Additionally, measurements can be performed at various x-ray edges which makes RIXS element and orbital sensitive, including being able to probe the local environment~\cite{ament2011resonant}. The two common edges typically probed include the $K$ $(1s\rightarrow 4p~\text{transition})$ and $L$ $(2p\rightarrow 3d~\text{transition})$-edge. Angular momentum conservation rules prohibit a single spin-flip excitation, but allow for a double spin-flip in the $K$-edge indirect RIXS process~\cite{van2007theory,PhysRevB.77.134428}. In a similar spirit, the presence of spin-orbit coupling (SOC) in the intermediate $2p$ state allows for a single-flip excitation in the direct RIXS process at the $L$-edge~\cite{PhysRevLett.103.117003}. 

Resonant inelastic x-ray scattering has the ability to track single and multi-spin flip magnetic dispersions, identify the ordering patterns of a strongly correlated material and frustrated quantum magnet, map the dynamical structure factor (DSF)~\cite{PhysRevB.103.224427}, detect the possibility of higher-order magnetic excitation terms, and unravel the effects of quantum correction~\cite{de2024resonant,wang2024magnon}. Till date, there have been theoretical efforts which have elucidated the $K$-~\cite{van2007theory,PhysRevB.77.134428} and $L$-edge~\cite{PhysRevLett.102.167401,PhysRevB.85.064421,PhysRevX.6.021020} RIXS mechanism within the context of local spin-flip. The direct $L_3$-edge RIXS process involves both non-spin-conserving (NSC, $\Delta S=1$) and spin-conserving (SC, $\Delta S=0$) channels~\cite{PhysRevB.103.L140409}. In addition to the above, both RIXS and INS spectra can display the presence of satellite intensity peaks above the single-magnon dispersion curve. This is an intriguing spectral signature, which presently has a host of competing explanations. According to one theory, the spectral weight softens and the peaks broaden as the multimagnon continuum spectrum arises from the bimagnon \cite{PhysRevLett.100.097001,PhysRevB.77.134428} or the three-magnon~\cite{10.21468/SciPostPhys.4.1.001,PhysRevLett.115.207202,PhysRevB.85.064421,Igarashi_2015_effect} excitation feature. However, another viewpoint suggests the possibility of deconfined spinon pairs due to the presence of damped spectral features~\cite{PhysRevLett.105.247001,dalla2015fractional,ghioldi2016rvb,PhysRevX.12.021041}. The third scenario is related to quantum entanglement effects arising from the electronic spin, which could account for the energy loss of the coherent magnon spectrum in the Brillouin Zone (BZ) boundary~\cite{kim2024quantum,PhysRevB.82.144407}. The goal of this article is to investigate this issue. 

Spin-flip excitation in RIXS can be computed within the ultrashort core-hole lifetime (UCL) expansion formalism~\cite{PhysRevB.75.115118}. Two local spin-flips can create a two-magnon continuum in the RIXS SC channel~\cite{PhysRevB.77.134428}. The two-magnon can form a bound state resulting in the formation of a bimagnon which is a coherent state of a two-magnon continuum~\cite{PhysRevLett.74.1867}. The $1/S$-interacting spin wave theory analysis has been applied to explain the origins of spin excitation spectrum in RIXS at both the $K$ and $L$-edge. The influence of the magnon-magnon interaction on RIXS can be calculated via the Dyson equation and ladder diagram approximation~\cite{nagao2007two,luo2014spectrum,xiong2017magnon,PhysRevB.100.054410}. Beyond the two-magnon, one can also have a three-magnon excitation. This can be considered to originate from two different sources: the $1/S$ expansion of a single spin-flip scattering operator~\cite{PhysRevB.85.064421,Igarashi_2015_effect} and the one-to-three magnon hybridization processes. These could lead to the renormalization of high-energy excitations~\cite{10.21468/SciPostPhys.4.1.001,PhysRevLett.115.207202}. Additionally, the three spin-flip operators~\cite{ament2010strong,pal2023theoretical} of the RIXS  NSC channel with $\Delta S=1$ can generate a three-magnon continuum. This raises several key questions: (a) which spin excitation mechanism underlies the satellite intensity of the single-magnon peak detected in RIXS? (b) how does it relate to the materials underlying spin-spin correlation and symmetry properties? (c) should it be explained by a multimagnon excitation or is there an alternative possibility?

To address the above questions one can pursue a perturbative interacting spin wave theory approach. Within this formalism one constructs the excitations from an ordered ground state that breaks $SU(2)$ symmetry. The local spin-flip disperses through the lattice to generate magnon excitations. However, the interacting spin wave theory formalism does not allow us to consider quantum entanglement effects and to elucidate its consequence on the RIXS spectrum. Thus, to avoid any short comings of accurately considering \emph{both} intrinsic quantum fluctuation and entanglement effects which do not violate $SU(2)$ symmetry, we utilize the Schwinger-boson mean-field theory (SBMFT). Within this approach, we introduce the bond spin-flip process as an alternative RIXS scheme. We propose that RIXS can induce fluctuations of the resonant valence bond (RVB). We introduce average RVB components which captures a bond spin-flip RIXS mechanism that acts in conjuction with the local site single spin-flip UCL RIXS mechanism ~\cite{ghioldi2016rvb,doi:10.1073/pnas.0703293104}. Here, the intermediate state spin shake-up process arises from core hole scattering~\cite{PhysRevB.77.134428,PhysRevX.6.021020}. 

The bond spin-flip mechanism is treated within SBMFT. This theory maintains the $SU(2)$ symmetry of the Hamiltonian. The magnetic state is conceptualized as a combination of an ordered antiferromagnetic background (which dominates) and embedded in it are fluctuating RVB states~\cite{doi:10.1073/pnas.0703293104}. Note, our theory focuses exclusively on the AF phase. We are not considering phase transitions from an AF state to a full RVB state. Since, the spin operators in SBMFT preserves $SU(2)$ symmetry and can be represented as entangled states of two flavors of bosons, each spin-flip will involve the variation of the occupation number of two bosons. An extended continuum beyond spin-wave excitation has been predicted within the SBMFT formalism for INS~\cite{ghioldi2016rvb,PhysRevB.105.224404}. This theory has explained the satellite intensity of a single-magnon peak, which is due to the linear superposition of the singlet and the triplet excitation contribution~\cite{ghioldi2016rvb}. However, the process to generate spin excitations in RIXS is fundamentally different from INS, especially due to the presence of core-hole in the intermediate state. Thus, it is important to investigate this issue, separately. Next, considering the length of the article, in the following paragraph we summarize the main results of the paper.

We investigate the origin of the satellite intensity features above the single-magnon dispersion in the $L_3$-edge RIXS spectra of a square-lattice antiferromagnet. Motivated by the effective $\tilde{J}_1$–$\tilde{J}_2$–$\tilde{J}_3$ Heisenberg spin model of cuprates, we investigate the implications of fusing the effects of spin fluctuations within the N\'eel order and RVB fluctuations in a square lattice geometry~\cite{doi:10.1073/pnas.0703293104}. To provide a systematic development of the theory, we first compute the RIXS spectra of this Hamiltonian using the $1/S$-interacting spin wave theory and then using SBMFT. Our theoretical analysis considers multi-spin-flip excitations, their possible coherent bound-state configurations, and the contribution of fluctuating RVB configurations to account for a diverse set of $L$-edge magnetic RIXS spectrum features. Within the context of interacting spin wave theory, we examine how the frustration parameters influence the single magnon dispersion and the RIXS spectra of the two-spin, the four-spin, and the six-spin correlations. The two-spin correlation RIXS spectra show sharp single-magnon peaks which is consistent with spin-wave dispersion, which are suppressed by $J_2,J_3$ but become more pronounced at the $M$ point as $J_c$ increases. The four-spin correlation RIXS spectra exhibit characteristic two-magnon features, which become softened and broadened upon incorporating magnon-magnon interactions. The six-spin correlation spectra exhibit broad features at high energies, reflecting contributions from three-magnon response. Within the SBMFT approach, we propose a RIXS mechanism involving fluctuating RVB bond spin-flipping processes. We investigate the mean-field two-spinon and four-spinon RIXS responses. Based on our calculations, we find that the spin correlation spectrum computed to a higher order UCL expansion of the magnon response can be captured via a mean-field spinon response in the first order of UCL. One of the most important conclusions of our investigation is the following. The satellite intensity feature can originate from the dual effects of both one-to-three magnon hybridization and condensed spinons arising from RVB fluctuations. While spin wave theory can capture the effects of magnon hybridization, the insight that condensed spinons may also have a visible experimental role to play can only be understood within the context of a SBMFT formalism. This conceptual viewpoint provides a physically appealing perspective on how nonlocal entangled spinon processes can be encoded in the complex multimagnon dynamics observed in RIXS experiments. Additionally, our findings highlight the role of quantum fluctuations, entanglement, and Higgs mechanism which shape the RIXS spectra beyond the conventional magnonic interpretations.
 
This article is organized as follows. In Sec.~\ref{sec:modelmethod}, we introduce our method. Sec.~\ref{subsec:EffectiveHeisenbergmodel} illustrates the effective version of the $J_1-J_2-J_3-J_c$ model which we redefine as the $\tilde{J}_1-\tilde{J}_2-\tilde{J}_3$ model. We then outline the $1/S$ interacting spin wave theory (the details of which are supplied in the Appendix as mentioned below). Section~\ref{subsec:sbmft} introduces the SBMFT  Hamiltonian and the corresponding ground state configuration. In Sec.~\ref{sec:rixssop} we introduce the RIXS processes and construct the RIXS scattering operators. In Sec.~\ref{sec:resdis} we state the RIXS results and discuss them. In Sec.~\ref{subsec:1mag} we analyze the single-magnon dispersion and the magnetic RIXS spectral weight. In Sec.~\ref{subsec:4scf}, we analyze the RIXS spectrum of the four-spin-correlation manifesting as the two-magnon (Sec.~\ref{subsubsec:2mrs}), the bimagnon (Sec.~\ref{subsubsec:bi}), and the mean-field two-spinon (Sec.~\ref{subsubsec:2spi}). In Sec.~\ref{subsec:6scf}, we analyze the six-spin-correlation RIXS spectra based on the three-magnon  (Sec.~\ref{subsubsec:3mag}) and the mean-field four-spinon description (Sec.~\ref{subsubsec:4spi}).  In Appendix \ref{app:LSWT+1/S}, we introduce the spin wave formalism and the expression of the $1/S$-interacting Hamiltonian. In Appendix \ref{app:sbmft}, we state the details of the SBMFT calculation. In Appendix \ref{app:rso}, we supply the explicit expressions of the multimagnon and the mean-field spinon RIXS operator and scattering matrix elements. In Appendix \ref{app:ri}, we provide the steps to derive the RIXS intensity formulae based on the Green's function approach. In the Supplemental material~[\onlinecite{supp}], we provide tables that summarize the acronyms, notations, and mathematical symbols used in the article.

\section{Model and Methods \label{sec:modelmethod}}
\subsection{Model and $1/S$-interacting spin wave theory\label{subsec:EffectiveHeisenbergmodel}} 
\begin{figure}
\centering
\includegraphics[width=85mm]{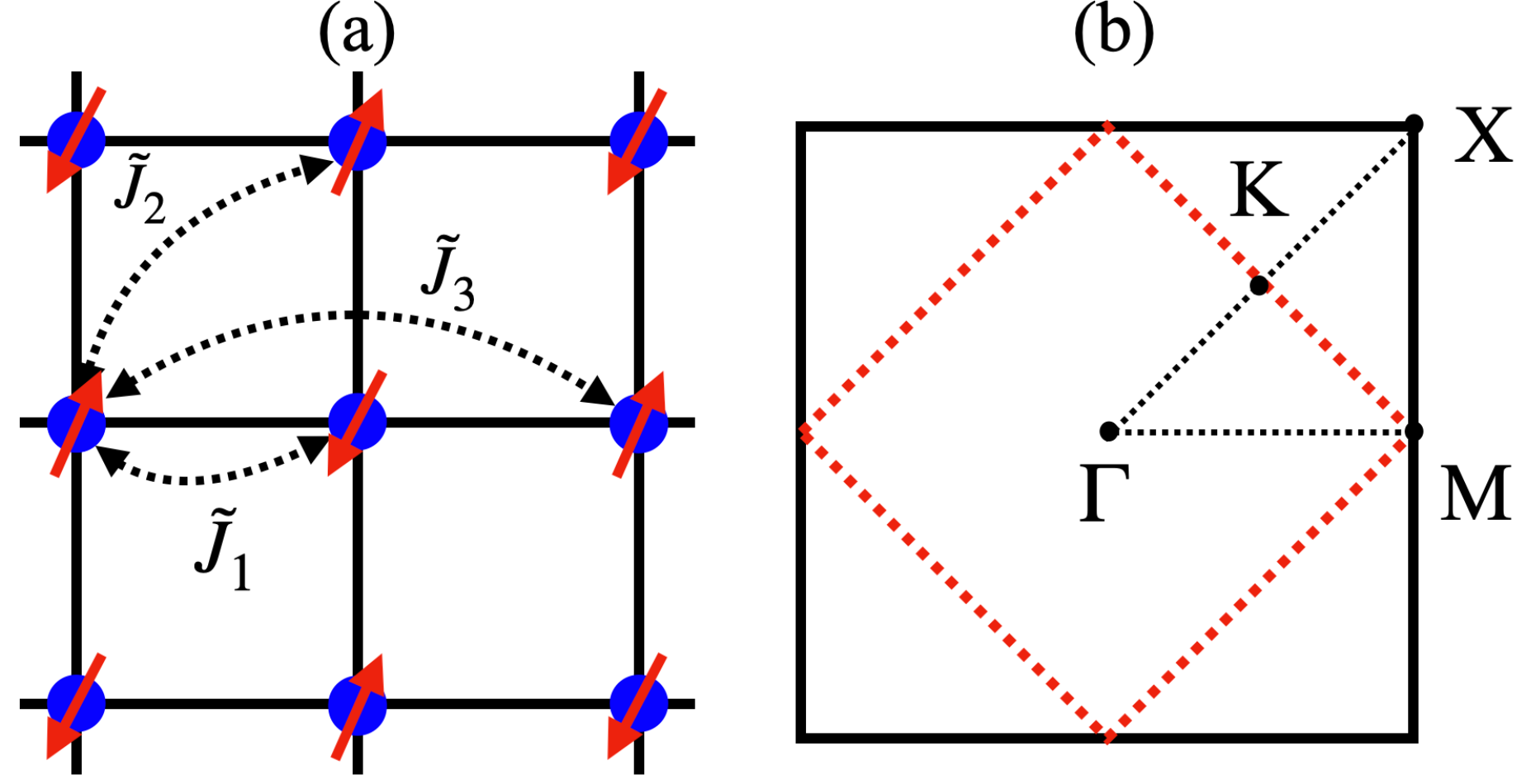}
\caption{ (a) Spin configuration of the N\'{e}el ordered state in a square lattice. Spin wave theory was constructed out of this broken $SU(2)$ symmetry configuration. The fluctuating RVB states considered in the Schwinger boson theory are not shown. The filled blue circles represent magnetic atoms. The red arrows denote the orientations of the spins. Antiferromagnetic ordering is exhibited as an example. Exchange interactions up to the third nearest neighbor are illustrated. The effective magnetic couplings $(\tilde{J}_1,\tilde{J}_2,\tilde{J}_3) = (J_1-2J_cS^2,J_2-J_cS^2,J_3) $ are defined in terms of the original exchange interactions where the first-neighbor is $J_1$, the second-nearest neighbor is $J_2$, the third-nearest neighbor is $J_3$, and the cyclic exchange coupling is $J_c$~\cite{PhysRevLett.86.5377,PhysRevB.79.235130}.  (b) The Brillouin zone of the square lattice. Red dotted line represents boundary of the magnetic Brillouin zone (MBZ), with high-symmetry points $\Gamma(0,0)$, $M(\pi,0)$, $K(\frac{\pi}{2}$, $\frac{\pi}{2})$, and $X(\pi,\pi)$.}
\label{fig:fig1}
\end{figure}

The low-energy magnetic excitation in cuprates can be modeled using an effective frustrated Heisenberg Hamiltonian with couplings $\tilde{J}_1$, $\tilde{J}_2$, and $\tilde{J}_3$, which incorporate the effects of cyclic four-spin interactions derived from a fourth-order $t/U$ expansion of the Hubbard model ~\cite{PhysRevLett.86.5377}. This effective model has been shown to successfully reproduce the spin-wave dispersions observed in neutron scattering and Raman experiments on cuprates \cite{PhysRevLett.86.5377,PhysRevB.39.2299}. One can arrive at the effective model by beginning with the frustrated $J_1-J_2-J_3-J_c$ Heisenberg model which is given by  \begin{align}
\label{eq:effham}
H &= J_1\sum\limits_{\langle i,j\rangle}\bs_i\cdot \bs_j 
    + J_2\sum\limits_{[i,j]}\bs_i\cdot \bs_j
    + J_3\sum\limits_{[[i,j]]}\bs_i\cdot \bs_j \nonumber\\
  &+ J_c\sum\limits_{i,j,k,l}\left\{(\bs_i\cdot \bs_j)(\bs_k\cdot \bs_l)
    + (\bs_i\cdot \bs_l)(\bs_k\cdot \bs_j)\right. \nonumber\\
  &\left. - (\bs_i\cdot \bs_k)(\bs_j\cdot \bs_l)\right\},
\end{align}
where $J_1>0$ is the nearest-neighbor AF interaction~\cite{PhysRevB.79.235130}. The second- and third- antiferromagnetic neighbor interactions are given by $J_2$ and $J_3$, respectively. The four-spin cyclic exchange interaction is given by $J_c$. The $J_1-J_2-J_3-J_c$ model can be recast into an effective $\tilde{J}_1-\tilde{J}_2-\tilde{J}_3$ form as illustrated in Fig.~\ref{fig:fig1}(a). Thus, we obtain the effective Hamiltonian as 
\begin{eqnarray}
\label{eq:eqH}
H&=&\tilde{J}_1\sum \limits_{i,\delta_1}\bs^A_i\cdot \bm{S}^B_{i+\delta_1}+ 
\tilde{J}_2\sum \limits_{i,\delta_{2}}\bs^A_i\cdot \bs^A_{i+\delta_{2}}+\bs^B_i\cdot \bs^B_{i+\delta_{2}}\nonumber\\
&+& \tilde{J}_3\sum \limits_{i,\delta_{3}}\bs^A_i\cdot \bs^A_{i+\delta_{3}}+\bs^B_i\cdot \bs^B_{i+\delta_{3}},
\label{efh}
\end{eqnarray}
where the effective frustration parameters are defined in the caption of Fig.~\ref{fig:fig1}. $A$ and $B$ represents the two sublattices. We note that Eq.~\eqref{efh} is analyzed using the Holstein-Primakoff (HP) boson representation under the assumption of a N\'{e}el ordered ground state. The details of spin wave theory and derivation are outlined in Appendix~\ref{app:LSWT+1/S}. 

The frustrated third neighbor model has successfully described magnetic excitations in the AF, the CAF and the valence bond crystal (VBC) phases of vanadates  and cuprates~\cite{PhysRevB.74.144422,PhilippeSindzingre_2010,PhysRevB.94.094438}. There are two stable phases: the classical AF N\'eel state with the ordering wave vector $\bq=(\pi,\pi)$, and the CAF stripe phases with the ordering wave vector $\bq=(\pi,0)$ and $\bq=(0,\pi)$, respectively. The remaining are spiral phases with valence bond crystal (VBC) features. The classical N\'{e}el state exists in the $0\leq J_2/J_1\leq 0.5$ and $0\leq J_3/J_1\leq 0.25$ region, while the CAF exists in the $0.5\leq J_2/J_1\leq 1$ and the $0\leq J_3/J_1\leq 0.25$ zone~\cite{PhysRevB.74.144422,PhilippeSindzingre_2010,PhysRevB.94.094438}. With other values of parameters that do not lie in the above specified ranges, the four-spin exchange interaction $J_c$ leads to a spin-nematic phase with a partial restoration of the $SU(2)$ symmetry. This occurs due to the interplay between frustration and quantum fluctuations~\cite{PhysRevLett.95.137206} which contributes to the formation of quartet binding~\cite{PhysRevLett.94.247004}. In our calculations, we limit the range of frustration parameters to $0\leq \tilde{J}_2/\tilde{J}_1\leq 0.1$, and $0\leq \tilde{J}_3/\tilde{J}_1\leq 0.1$, thereby, focusing only on the effects of the AF phase and its consequences on the $L_3$-edge RIXS excitation spectrum. 
 
We follow the standard procedure and apply the Holstein-Primakoff (HP) transformation, Fourier transformation, and Bogoliubov transformation (see Appendix.~\ref{app:LSWT+1/S} for details). We obtain the bosonic Hamiltonian of the $\tilde{J}_1-\tilde{J}_2-\tilde{J}_3 $ model as
\begin{equation}
H=\frac{-S^2Nz}{2}(\tilde{J}_1-\tilde{J}_2-\tilde{J}_3)+H_0+H_1.
\end{equation} 
The first term corresponds to the classical energy of the ground state, where $N$ is the number of lattice sites and $z=4$ is the coordination number. $H_0$ is the $\mathcal{O}(S)$ linear spin wave theory term of the spin wave expansion. The last term $H_1$ is the $\mathcal{O}(1/S)$ correction to the Hamiltonian. This includes both the Oguchi correction and the quartic spin-wave interactions. Based on the spin wave perturbation expansion scheme, the renormalized magnon dispersion can be written as
\begin{equation}
\omega_\kk=4J_1S\left(\kappa_\kk\epsilon_\kk+\frac{A_\kk}{2S}\right),
\label{eq:wk}
\end{equation}
where the detailed procedure to derive the interacting spin wave theory formulae and the expressions for $\kappa_\kk,\epsilon_\kk$, and $A_\kk$ are stated in Appendix~\ref{app:LSWT+1/S}.

We note that interacting spin wave theory formalism is a reliable theoretical tool to describe the low-energy excitations of a quantum magnet where a preferred $S^z$ direction is selected \cite{pires2021theoretical}. While the approach has been quite successful in predicting non-trivial features of the multimagnon RIXS spectrum~\cite{luo2014spectrum,xiong2017magnon,PhysRevB.100.054410}, one encounters technical issues when applying it to the case of the interacting three-spin flip case scenario. Based on our calculations, we find that the three-body interaction term cannot be treated in a controlled (analytical) manner with the Dyson and Faddeev equations to obtain the correct excitation spectrum of the three-magnon bound state. The perturbative summation of the interaction kernel leads to the emergence of Faddeev spurious states~\cite{PhysRevC.63.034313}. Thus, we conclude that the coherency of the three-magnon continuum is not appropriately captured by the $\mathcal{O}(1/S)$ spin wave theory and Dyson equation. While the continuous unitary transformation (CUT) method can calculate the three-body interaction in spin systems~\cite{schmiedinghoff2022three}, there are some known limitations of the CUT approach~ \cite{dusuel2004quartic,drescher2011truncation}. In the next section, we pursue the SBMFT approach that avoids the above technical deficiencies. 

\subsection{Schwinger-boson mean-field theory}
\label{subsec:sbmft}
Schwinger boson mean field theory offers an alternative perspective to spin wave theory for evaluating quantum fluctuations in magnetic systems. Since spin wave theory assumes a ground state with broken SU(2) symmetry, it cannot fully capture the effects of large quantum fluctuations and quantum entanglement~\cite{auerbach2012interacting}, which are crucial in several quantum magnets. However, SBMFT preserves SU(2) symmetry, and addresses these limitations by rewriting spins in terms of spinons (bosonic in our case). Subsequently, the Heisenberg interaction is expressed in terms of singlet bond operator $\hat{A}_{ij}$ (defined below). The spinons could be deconfined, confined, or condensed. In our case we are dealing with the later two situations where the spinons are confined in pairs due to gauge-mediated interactions with a confining potential or condensed in the Higgs phase~\cite{PhysRevB.49.4368,PhysRevB.52.440}. Note, we are not dealing with the case of spinons arising from fractionalization~\cite{lacroix2011introduction}. Through the mean-field order parameters, SBMFT explores both low-energy excitations and high-energy renormalization, revealing features like the continuum spectrum beyond spin-wave excitations~\cite{ghioldi2016rvb,PhysRevB.105.224404}. This framework highlights the role of tightly bound spinon pairs in shaping magnonic spectra, especially near the Goldstone mode region.

The spinon operators $a^{s}_i$ and $b^{s}_i$ are defined as $a^{s\dagger}_i|0\rangle_{\mathrm{SB}} =|\uparrow_i\rangle$ and $b^{s\dagger}_i|0\rangle_{\mathrm{SB}}=|\downarrow_i\rangle$, where $|0\rangle_{\mathrm{SB}}$ is the Schwinger-boson vacuum. The $|\uparrow_i\rangle$ and $|\downarrow_i\rangle$ represents $S=\frac{1}{2}$ states on lattice site $i$. We introduce spin operators in SBMFT as $\bs_i^p=\frac{1}{2}\varphi_i^\dagger \sigma_p\varphi_i$, 
where $\varphi_i^\dagger=(a_i^{s\dagger}, b_i^{s\dagger})$ is the spinor and $\sigma_p$ ($p=x,y,z$) is the Pauli matrix~\cite{PhysRevLett.61.617,auerbach2012interacting}. This representation leads to spin-flip processes that change the occupation number of the two bosons.  For simplicity, we only consider the $J_1$ interaction in the square lattice SBMFT. Note, in order for interacting spin wave theory to accurately capture the full effects of frustration, one must include all the relevant exchange interactions and consider a spin wave theory well beyond the linear level. However, as shown later in this article, interacting spin wave theory formalism is not stable to allow us to explain the origins of the satellite intensity peak above the single magnon peak, please see the last paragraph in Sec.~\ref{subsubsec:3mag}. Additionally, the simultaneous computation of a higher order interacting spin-wave theory with a higher order UCL expansion is a tedious and cumbersome endeavor. Thus, we attempted the SBMFT reformulation of the single-site UCL RIXS formalism for a minimal model that can capture the needed physics. Furthermore, at a practical level, inclusion of $J_2$ and $J_3$, leads to complicated self-consistent equations, that do not necessarily provide any further insight into the physics of the RIXS spectrum and the interpretation of the satellite intensity peak, especially when the spinons are confined and condensed (the phase we are interested in from an experimental perspective). Even at this level of simplification, our calculation of the SBMFT RIXS spectra indicates the presence of non-trivial features. 

We define a singlet bond operator $\hat{A}_{ij}=\frac{1}{2}[a^s_ia^s_j+b^s_ib^s_j]$ to represent antiferromagnetic correlation~\cite{PhysRevB.47.12329}. Using the definitions of the previous paragraph the SBMFT Hamiltonian is given by $\hat{H}=J_1\sum\limits_{\langle i,j\rangle}\left(S^2-2\hat{A}^\dagger_{ij}\hat{A}_{ij}\right)$.  By adding a local constraint $\lambda$ within the mean-field decoupling Ansatz, and applying the Bogoliubov transformation, we obtain the mean-field spinon dispersion as $\omega^s_\kk=\sqrt{\lambda-z\lvert A_{\delta}\gamma^A_\kk\rvert^2}$, where $\langle\hat{A}_{ij}\rangle=\langle\hat{A}_{ij}^\dagger\rangle=A_\delta$ is the mean-field order parameter. The details of these calculations are outlined in Appendix~\ref{app:sbmft}. The dispersion relationship $\omega^s_\kk$ of the mean-field spinon shows an exponentially small gap when $T\rightarrow 0$, a result which is consistent with the $1/S$ corrected interacting spin wave theory formalism~\cite{PhysRevLett.61.617,PRXQuantum.4.030332,PhysRevB.40.5028}. 

By minimizing the free energy expression Eq.~\eqref{eq:free_F} stated in Appendix~\ref{app:sbmft}, we obtain the self-consistent equations to identify the mean-field parameters $\{\lambda, A_{\delta}\}$. These are given by
\begin{subequations} 
\label{eq:spinmf} 
\begin{eqnarray} &\frac{1}{2N}\sum\limits_\kk\frac{ \lambda}{\omega^s_\kk}=\left(S+\frac{1}{2}\right),&\\ &\frac{1}{N}\sum\limits_\kk\frac{ \gamma_\kk^A}{\omega^s_\kk}\sin(\kk\cdot \delta)= A_{\delta}.&
\end{eqnarray} 
\end{subequations} In the thermodynamic limit $N\rightarrow \infty$ and $T\rightarrow 0$, spinons emerge as a Bose-Einstein condensate (BEC) at $\pm\frac{\bq}{2}$, where $\bold{Q}=(\pi,\pi)$ is the AF ordering wave vector ($\omega^s_{\pm\frac{\bq}{2}}\sim 0$). The mean-field order parameter $A_{\delta}=0.579$ and magnetization $m_0=0.3034$.~\cite{auerbach2012interacting,PhysRevB.47.12329,PRXQuantum.4.030332}. The Schwinger-boson ground state $|\mathcal{GS}\rangle_{\mathrm{SB}}$ is a product of the singular state $|g\rangle$ and the continuum state $|c\rangle$: \(|\mathcal{GS}\rangle_{\mathrm{SB}}=|g\rangle|c\rangle\) ~\cite{chandra1990quantum}. The singular state $|g\rangle$ represents the quantum corrected N\'eel state
\begin{flalign}
     &|g\rangle=\exp\left[\sqrt{\frac{Nm_0}{2}}\left(\alpha^{s\dagger}_{\frac{\bold{Q}}{2}}+\alpha^{s\dagger}_{-\frac{\bold{Q}}{2}}+i\beta^{s\dagger}_{\frac{\bold{Q}}{2}}+i\beta^{s\dagger}_{-\frac{\bold{Q}}{2}}\right) \right]|0\rangle_{\mathrm{SB}},&
     \label{gssb:s}
\end{flalign} 
where $\alpha^s$ and $\beta^s$ denote the Bogoliubov spinon operators. The continuum part $|c\rangle$ is the isotropic normal fluid state representing zero-point quantum fluctuations. It is given by the expression
\begin{flalign}
&|c\rangle=\exp\left[\sum_{\kk,\kk\neq\pm\frac{\bold{Q}}{2}} \mathrm{m}_\kk \alpha^{s\dagger}_\kk \beta^{s\dagger}_{-\kk}\right]|0\rangle_{\mathrm{SB}}, &      
\label{gssb:c}
\end{flalign} 
where $\mathrm{m}_\kk=v^s_\kk/u^s_\kk$ and $(u^s_\kk, v^s_\kk)$ represents the Bogoliubov coefficients (see Appendix.~\ref{app:sbmft}). The Schwinger-boson ground state can be linked to RVB states though gauge transformations~\cite{chandra1990quantum}. However, note this is not a true RVB state because the constraint is only satisfied on average~\cite{ghioldi2016rvb}. We thus treat it as an averaged RVB component, which is defined in the real space as
\begin{flalign}
&|c\rangle\approx\exp\left[\sum_{i,j} \mathrm{m}_{ij} \hat{A}^\dagger_{ij} \right]|0\rangle_{\mathrm{SB}}. & 
\label{gssb:crvb}
\end{flalign} 
In the above $\mathrm{m}_{ij}$ is the Fourier transform of $\mathrm{m}_\kk$, which can be conceptualized as the pairing amplitude of this averaged RVB component~\cite{fradkin2013field}. The infinite square lattice ground state can be viewed as a series of $1/S$-corrected N\'eel state with fluctuating RVB configurations~\cite{doi:10.1073/pnas.0703293104}. Note, to keep the calculation tractable, we consider only the isotropic singlet bonds which show up at the $J_1$ interaction level within the SBMFT model. Furthermore, the SBMFT Hamiltonian is quartic, so its results intrinsically include magnon-magnon interaction effects~\cite{PhysRevLett.61.617}. As we will show later, the bond spin-flip mechanism developed using the Schwinger boson approach is able to capture the quantum entanglement effects that are intrinsic to the material and can explain the satellite intensity of the single-magnon dispersion.

Before we end this section, we would like to emphasize a few subtle conceptual details. First, we are not double counting the physical implications of quantum fluctuations by considering both interacting spin wave theory and the Schwinger boson approach. Instead, they represent distinct theoretical treatments~\cite{doi:10.1073/pnas.0703293104}: the interacting spin wave formalism captures semiclassical spin-wave excitations around a N\'{e}el ordered state, whereas the Schwinger boson framework describes RVB-like quantum fluctuations. By analyzing the effects of these fluctuations simultaneously, we aim to explore how different aspects of magnetic excitations emerge under contrasting physical assumptions -- namely, long-range Néel order captured by spin wave theory versus a spin disordered background encoded in the RVB mean-field treatment. Second, from a physical perspective, the confined spinon corresponds to excitations in the neighborhood of the Goldstone mode~\cite{mezio2012low,PhysRevB.105.224404}, providing access to the RIXS spectrum beyond conventional spin-wave theory, while the condensed spinon represents the Higgs phase (the phase we are interested in from an experimental viewpoint)~\cite{PhysRevB.98.184403}. To describe the satellite peak, it is therefore sufficient to capture the spin dynamics of confined and condensed spinons within the six-spin correlation function probed by RIXS. We note that deconfined spinon phases associated with frustrated $Z_2$ ansatz are certainly interesting, but they lie beyond the scope of the present work, which is restricted to understanding the satellite intensity in the Néel-ordered (Higgs) phase. 

\section{RIXS process}\label{sec:rixssop}
In this section, we introduce the RIXS process for multimagnon and multispinon excitations within the $1/S$ interacting spin wave theory formalism, as well as the bond spin-flip scheme  based on the SBMFT formalism. Before proceeding with our derivation and discussion, we will first clarify our choice of the initial and the final states. In the ordered magnetic phase the initial and the final states are given by $|\kk_{in},\epsilon\rangle$ and  $|\kk_{out},\epsilon'\rangle$, where $\kk_{in} (\kk_{out})$ is the incident (outgoing) photon momentum. The incident (scattered) photon polarization is given by $\epsilon(\epsilon')$. The momentum difference $\qq$ between the outgoing and the incident photon is $\qq=\kk_{out}-\kk_{in}$. For the SBMFT analysis, we take the initial and the final states of the RIXS process as $|g\rangle \otimes |c\rangle$ and $|m\rangle \otimes |s\rangle$. Here, $|g\rangle$ represents the quantum-corrected N\'{e}el state, $|c\rangle$ is the averaged RVB state (which is similar to the continuum part mentioned in the previous section), $|m\rangle$ is the multimagnon eigenstate with energy $\omega_{n\qq}$ with respect to $|g\rangle$ and $|s\rangle$ is the mean-field spinon eigenstate with energy $\omega_s$ with respect to $|0\rangle_{\mathrm{SB}}$. For example, $\omega_{1\qq}$ is for single-magnon, $\omega_{2\qq}$ for two-magnon, etc.   The mean-field two (four) -spinon energies are given by $\omega_{2s}$ ($\omega_{4s}$).

\subsection{Local spin-flip RIXS scheme and spin wave RIXS matrix element}
\begin{figure*}
\includegraphics[width=\textwidth]{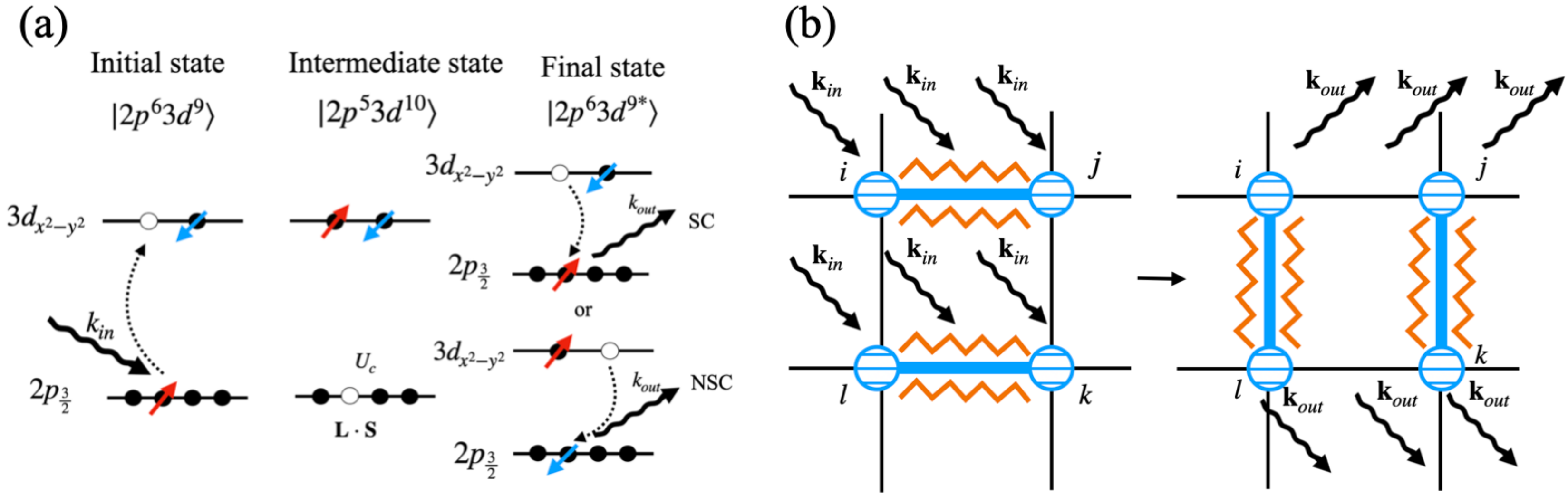} 
\caption{(a) Single site $L_3$-edge RIXS process in the spin-conserving (SC) and non-spin-conserving (NSC) channels. In the initial state, an incident photon excites a single electron to the $3d_{x^2-y^2}$ orbital leaving behind a core-hole in the $2p$ orbital. The excited electron fills the empty hole in the $3d$ orbital. In the intermediate state, there are effects of the core-hole potential and spin-orbit coupling. In the final state the electron loses its energy and falls back to the $2p$ orbital. If it has the same spin orientation it belongs to the SC channel. If has the opposite spin orientation it belongs to the NSC channel. (b) $L_3$-edge bond spin-flip RIXS mechanism. The solid blue lines represent singlet bonds. The blue hatched circles represent sites where a single-spin flip mechanism is active. The wavy orange lines represent gauge interaction (the glue that holds the bond together). Incident photons can induce resonating valence bonds fluctuations. Initially photons strike an existing RVB bond (panel(b), left figure) and ruptures it. Next, the decoupled sites undergo a core-hole mediated RIXS spin-flip at each site (within the UCL scheme). After, the single-site process is completed, the photons leave the system. However, the RVB bonds can be restored in another orientation (or not) due to the local constraint $\lambda$ and due to intrinsic parity-gauge in the Schwinger-boson ground state~\cite{fradkin2013field,chandra1990quantum}. If the bonds are reorientated then a RIXS photon induced RVB fluctuation is created. Such a bond spin-flip mechanism is able to capture a three-site spin correlation (described within a bond spin-flip process). Based on our RIXS intensity calculations, we show that there is a finite amplitude of such a process occurring in a quantum magnet.}
    \label{fig:fig2}
\end{figure*}

In the direct $L_3$-edge RIXS process, the photon induces a local spin-flip, leading to single and multimagnon excitations. Quantum fluctuations can enable simultaneous excitation of multiple magnons. First, we review the perturbation theory of RIXS in an insulating quantum magnet. Two expansion methods can be implemented simultaneously to capture the effects of x-ray scattering and spin wave quantum fluctuations. The UCL expansion captures spin-spin correlations in the x-ray scattering process~\cite{PhysRevB.75.115118,PhysRevB.77.134428,PhysRevB.106.L060406}. Spin-wave expansion, $\mathcal{O}(1/S)$ in our case, accounts for quantum fluctuations of spin excitation~\cite{canali1992theory,PhysRevB.82.144402}. We denote the $\mathcal{L}$-th order of the UCL expansion as $\mathcal{O}(\text{UCL}[\mathcal{L}])$. Similar to the spin wave expansion of the Hamiltonian, which can be expressed in order $S^{2-n/2}$~\cite{PhysRevB.82.144402}, the RIXS scattering operator can also be spin wave expanded in the order $\mathcal{M}=S^{m-n/2}$, where $m$ is the number of spin operators in the RIXS operator and $n$ is the number of HP bosons. In the following, we will write the general expression of RIXS operator matrix element in the format $\mathrm{O}^{(\mathcal{L},\mathcal{M},\mathcal{N})}_{\qq}$ where $\mathcal{N}$ is the order of single spin-flip $\hat{S}^x_i$ operator where $\mathcal{N}=0$ ($\mathcal{N}=1$) represents the SC (NSC) channels. Therefore, a general UCL and spin wave perturbation expanded RIXS operator matrix element can be written as~\cite{PhysRevB.77.134428,ament2010strong,PhysRevB.103.L140409}
\begin{align}
\mathrm{O}^{(\mathcal{L},\mathcal{M},\mathcal{N})}_{\qq}=\textstyle{\langle \mathrm{n}|}\sum\limits_{i(,j,l)}e^{i\qq r_i}(\hat{S}^x_i)^\mathcal{N} h_i(\hat{\bs}_i\cdot \hat{\bs}_j)^\mathcal{L} h^\dagger_i|\mathcal{GS}\rangle_{\textrm{swt}},
\label{eq:oql}
\end{align}
where $h^\dagger_i$ is the core-hole creation operator, $|\mathcal{GS}\rangle_{\textrm{swt}}$ is the spin wave theory N\'{e}el state, and $|\mathrm{n}\rangle$ is the excited state of the HP bosons. $(\hat{\bs}_i\cdot \hat{\bs}_j)^\mathcal{L}$ is the spin-spin correlation term between lattice points. $(\hat{\bs}_i\cdot \hat{\bs}_j)^\mathcal{L} \in \{1, (\hat{\bs}_i\cdot \hat{\bs}_j), (\hat{\bs}_i\cdot \hat{\bs}_j)(\hat{\bs}_i\cdot \hat{\bs}_l)$\} for $\mathcal{L}=0,1,2$, respectively with $j,l$ the nearest-neighbors of $i$ ($j\neq l$). The calculation details are supplied in Appendix~\ref{app:LSWT+1/Sso}. The RIXS intensity $I^{(\mathcal{L},\mathcal{M}^2,\mathcal{N})}(\qq,\omega)$ is computed using the expression \begin{eqnarray}
    I^{(\mathcal{L},\mathcal{M}^2,\mathcal{N})}(\qq,\omega)\hspace{-0.05cm} &\propto& \hspace{-0.1cm} \sum\limits_{n}\mathrm{O}^{\dagger(\mathcal{L},\mathcal{M},\mathcal{N})}_{\qq}(t)\mathrm{O}^{(\mathcal{L},\mathcal{M},\mathcal{N})}_{\qq}(0) \delta(\omega-\omega_{n\qq})\nonumber\\
    &=&-\frac{1}{\pi}\mathrm{Im}[G(\qq,\omega)],
    \label{eq:stgreen}
\end{eqnarray}
where $\omega_{n\qq}=\omega_{\kk_1}+\cdots+\omega_{\kk_n}$ is the multimagnon excitation energy with $\sum\limits_i^n \kk_i=\qq$. The UCL expansion of the RIXS cross-section can be mapped both to the SC and the NSC channels~\cite{PhysRevX.6.021020,PhysRevB.106.L060406,pal2023theoretical}. Following the derivation details outlined in the Appendix \ref{app:ri}, the multimagnon RIXS intensity $\mathcal{I}_{mm}(\qq, \omega)$ up to $\mathcal{O}(\text{UCL}[2])$ is given by 
\begin{widetext}
\begin{eqnarray}
\mathcal{I}_{mm}(\qq, \omega) &\propto& 
\frac{1}{\Gamma^2} \left[I^{(0,S,1)}_{\mathrm{1m}}+\xi^2 I^{(1,S^2,0)}_{\mathrm{2m}}  +\xi^4 I^{(2,S^2,0)}_{\mathrm{2m}}+I^{(0,1/S,1)}_{\mathrm{3m}} + \xi^2 I^{(1,S^3,1)}_{\mathrm{3m}}  +\xi^4 I^{(2,S^3,1)}_{\mathrm{3m}}\right].
\label{eq:rixsiex}
\end{eqnarray}
\end{widetext}
The subscripts $\mathrm{1m}$, $\mathrm{2m}$, and $\mathrm{3m}$ denote single-, two-, and three-magnon, respectively. The derivation of the above equation is based on the single-site RIXS mechanism sketched in Fig.~\ref{fig:fig2}(a). Note, we retain only the leading terms with the coefficient $\xi=\tilde{J}_1/\Gamma=0.375$, where $\Gamma$ is the inverse core-hole lifetime \cite{PhysRevX.6.021020}. The subleading terms associated with $\tilde{J}_2$ and $\tilde{J}_3$ are neglected, as their squared coefficients, $|\tilde{J}_2/\Gamma|^2 \approx 0.031^2$ and $|\tilde{J}_3/\Gamma|^2 \approx 0.008^2$ are two orders of magnitude smaller than $\xi^2$. The corresponding multimagnon RIXS spectral weight $ W_{nm}^{{(\mathcal{L},\mathcal{M}^2,\mathcal{N})}}(\qq)$ is computed from the expression \begin{equation}
W_{nm}^{{(\mathcal{L},\mathcal{M}^2,\mathcal{N})}}(\qq)=\int d\omega I_{nm}^{{(\mathcal{L},\mathcal{M}^2,\mathcal{N})}}(\qq,\omega).
\label{eq:weight}
\end{equation}

\subsection{Bond spin-flip scheme and SBMFT RIXS matrix element}\label{subsec:sbmftrixsso}
The UCL expansion in RIXS is based on a local spin-flip scheme~\cite{PhysRevB.75.115118}. However, we propose that due to the presence of an intermediate state manifold, which permits a spin shake-up at the $L$-edge, a combination of \emph{both} local spin-flip and bond spin-flip can occur during the RIXS scattering process. Thus, there will be RIXS induced RVB fluctuations arising from a bond spin-flip process as illustrated in Fig.~\ref{fig:fig2}(b). To mimic the bond spin-flip mechanism showing the results of preserving the SU(2) symmetry, we map the RIXS scattering operators to a Schwinger boson representation. Implementing a mean-field decoupling [Eq.~\eqref{mfa}] using the mean-field ansatz $\{
\lambda, A_{\delta}\}$, and performing the Bogoliubov transformation, we can map the two-, three- spin-flip RIXS scattering operators to the mean-field two-, four-spinon RIXS response. In our calculation, $\hat{O}^{SB}_{\qq,\mathrm{2s}}$ and $\hat{O}^{SB}_{\qq,\mathrm{4s}}$ represents the mean-field two-, four- spinon RIXS scattering operators, respectively. The SBMFT two-spinon RIXS scattering matrix element that originates from a two-spin flip RIXS operator matrix element is given by 
 \begin{align}
\mathrm{O}^{SB}_{\qq,\mathrm{2s}} \propto \langle \mathrm{2s} | \sum\limits_{i,j} e^{i\qq \cdot \br_i}  h_i(\hat{\bs}_i\cdot \hat{\bs}_j)_{MF} h^\dagger_i |\mathcal{GS}\rangle_{\mathrm{SB}},
    \label{eq:o2smf}
\end{align}
where $|\mathrm{2s}\rangle$ is the mean-field two-spinon excited state. Similarly, the three-spin-flip RIXS scattering operator generates the mean-field four-spinon RIXS operator matrix element as
\begin{align}
O_{\qq,\mathrm{4s}}^{SB}&\propto\langle \mathrm{4s}|\sum\limits_{i,j,l}e^{i\qq \cdot \br_i}\hat{S}^x_ih_i(\hat{\bs}_i\cdot \hat{\bs}_j)_{MF} h^\dagger_i|\mathcal{GS}\rangle_{\mathrm{SB}},
 \label{eq:o4s}
\end{align} where $|\mathrm{4s}\rangle$ is the excited state of the mean-field four-spinon.

In the above scattering operators $(\hat{\bs}_i\cdot \hat{\bs}_j)_{MF}$ is the spin correlation expressed in terms of the singlet bond operator $\hat{A}_{ij}$ within the mean-field Ansatz. Its spin dynamics are governed by the mean-field Hamiltonian $\hat{H}_{MF}$, Eq.~\eqref{eq:hmf}. To go beyond the mean-field level,  one needs to consider gauge fluctuations around the mean-field order parameter $A_{\delta}$. The nature of these gauge fluctuations can be addressed by the Invariant Gauge Group (IGG) formalism~\cite{PhysRevB.65.165113,jacobsen2010exact,lacroix2011introduction}. In this viewpoint, when RIXS measures spin-spin correlation within the mean-field ansatz at $T\to 0$, the massless gauge fluctuation can confine the spinons in tightly bound pairs. The confined spinons will behave as magnonic-like excitations that will contribute to the continuum part of the Schwinger boson contribution~\cite{PhysRevB.49.4368,PhysRevB.52.440,PhysRevB.50.16428,ghioldi2016rvb}. Additionally, the BEC of spinons at $\pm \bq/2$ point can drive the system to a Higgs phase, referred to as the singular part~\cite{PhysRevB.98.184403,PhysRevB.73.075119}. At this condensation point, the gauge bosons become massive via the Higgs mechanism, a phenomenon that we show can be detected by RIXS spectroscopy~\cite{PhysRevB.73.075119,PhysRevB.42.4568,kim1999theory,PhysRevB.53.R14729}. 

We will demonstrate in later sections, that the spin correlation spectra obtained by summing up all contributions up to $\mathcal{O}(\text{UCL}[2])$ using the local spin-flip scheme (within spin wave theory) is reproduced by the continuum part of the spectra predicted using $\mathcal{O}(\text{UCL}[1])$ scattering operators within SBMFT framework. In other words, we conclude that the higher-order spin correlation spectra predicted by an UCL expansion within the local spin-flip scheme can be effectively captured by RVB fluctuations within the bond spin-flip scheme. The Schwinger-boson formalism yields physical information which is not readily accessible from the $SU(2)$ symmetry broken spin wave theory approach. The singular part of the four-spin correlation spectra captures the sharp peaks in both the continuum part spectrum and the two-magnon spectrum. The singular part contribution of the six-spin correlation function exhibits the satellite intensity peaks above the single magnon dispersion in the Higgs phase, see discussion in Sec.~\ref{subsec:6scf}. 
 
\section{Results and discussion\label{sec:resdis}}
In this section, we present the calculated RIXS spectra in the following order: two-spin correlation, four-spin correlation, and six-spin correlation. Note, we consider the effects of additional interactions $J_2,J_3$, and $J_c$ which induce relatively small but noticeable spectroscopic signatures in the undoped cuprates~\cite{PhysRevLett.86.5377,PhysRevB.79.235130,PhysRevB.39.2299}. The two-spin correlation is computed within the $1/S$ interacting spin wave theory formalism, only. The last two correlations are calculated both with the $1/S$ interacting spin wave theory and the SBMFT approach. Furthermore, to avoid cluttering the section we have stated the actual expressions of the RIXS scattering operators in the Appendix. For convenience, in this section, we simply refer to the equation numbers from Appendix \ref{app:msme} and \ref{app:ssme} that are used to derived the RIXS intensity expressions.

\subsection{Two-spin correlation}\label{subsec:1mag}
\begin{figure*}[htbp]
	\centering
\includegraphics[width=\linewidth,scale=1.00]{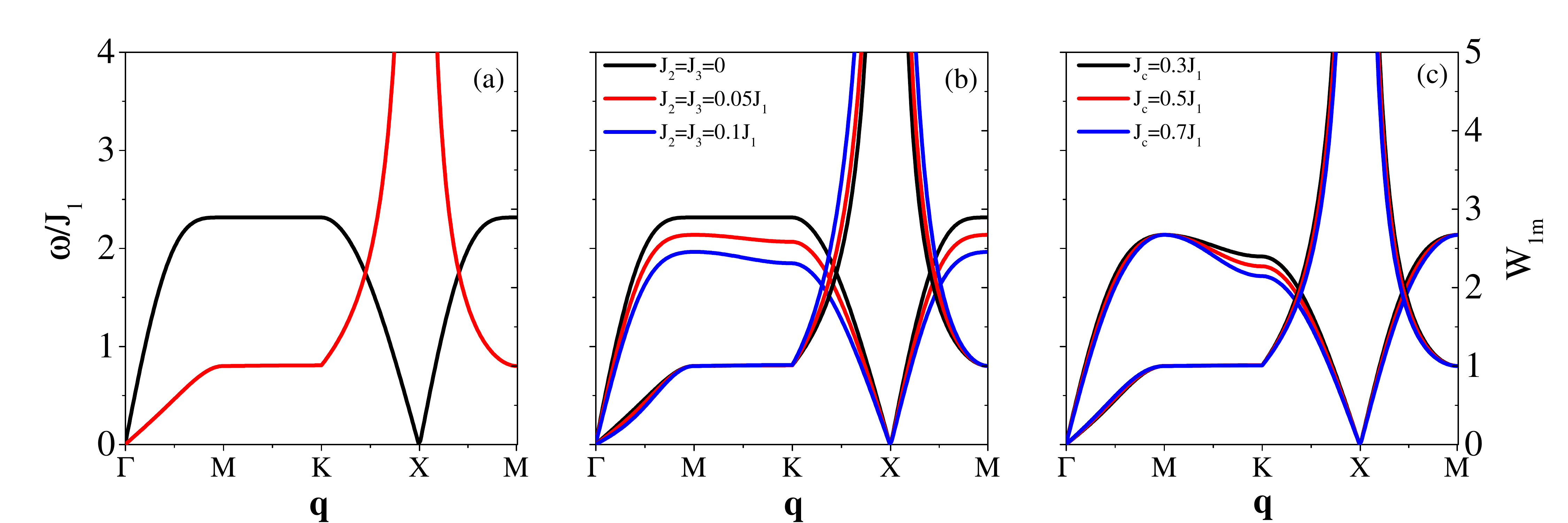}
\caption{Renormalized single magnon dispersion $\omega_\qq$ (the left axis) and $1/S$ -corrected $L_3$-edge (single magnon) RIXS spectral weight $W_{\mathrm{1m}}$ (the right axis). The momentum $\qq$ path is $\mathrm{\Gamma}-\mathrm{M}-\mathrm{K}-\mathrm{X}-\mathrm{M}$. (a) $(J_2,J_3,J_c)=0$. Along the MBZ boundary path $\mathrm{M}-\mathrm{K}$ the magnon band is flat with $\omega \sim 2.3J_1$. At the $\mathrm{X}$ point, the dispersion is gapless while the single-magnon spectral weight diverges~\cite{PhysRevLett.105.167404}. (b) $J_2=J_3=(0,0.05,0.1)J_1$ and $J_c=0$. As the frustration parameters $(J_2, J_3)$ increase, the dispersion energy is suppressed. However, the single-magnon RIXS spectral weight $W_{\mathrm{1m}}$ is relatively insensitive to the changes. (c) $J_c=(0.3,0.5,0.7)J_1$ and $J_2=J_3=0.05J_1$. The cyclic exchange interaction $J_c$ predominantly reduces the magnon energy at the $\mathrm{K}$ point, but has minimal influence on $W_{\mathrm{1m}}$. }
\label{fig:fig4}
\end{figure*} 
The two-spin correlation function of the RIXS intensity Eq.~\eqref{eq:stgreen} involves two spin operators. Consequently, we restrict the calculation to the case where $\mathcal{L}=0$ and $\mathcal{N}=1$ in Eq.~\eqref{eq:stgreen} with the RIXS scattering operators given by Eq.~\eqref{eq:o0s1m} and Eq.~\eqref{eq:o03m}. This yields the single- and the three- magnon spectra $I^{(0,S,1)}_{\mathrm{1m}}$ and $I^{(0,1/S,1)}_{\mathrm{3m}}$, respectively. Within a local spin-flip scenario, the two-spin correlation function manifests as a single magnon RIXS response. The single magnon direct RIXS process at the $L_3$-edge is facilitated by SOC in the intermediate state, created by the annihilation of a core-hole~\cite{PhysRevLett.103.117003,PhysRevLett.102.167401,ament2010strong}. The Hamiltonian also includes a one-to-three magnon hybridization process at $\mathcal{O}(1/S)$ which contributes to the spectroscopic behavior of the two-spin correlation function~\cite{PhysRevB.72.014403,PhysRevB.85.064421,Igarashi_2015_effect,PhysRevLett.115.207202,10.21468/SciPostPhys.4.1.001}.

\subsubsection{Single-magnon dispersion curve}

Figure.~\ref{fig:fig4} shows the renormalized magnon dispersion [Eq.~\eqref{eq:wk}] and single-magnon RIXS spectral weight [Eq.~\eqref{eq:weight}] calculated using the RIXS scattering operator specified by Eq.~\eqref{eq:o0s1m}. The renormalized single-magnon dispersion exhibits a flat band along the MBZ boundary path $\mathrm{M}-\mathrm{K}$ for the unfrustrated Heisenberg model with $J_1$ interaction only, see Fig.~\ref{fig:fig4}(a). This behavior is a result of creating a single magnon from a local spin-flip, which has an excitation energy of $2zS^2J_1=2J_1$ within linear spin wave theory~\cite{verresen2018quantum}. The $1/S$ Oguchi correction contributes an additional $\sim 0.3 J_1$ to the total energy. Notably, at the $\mathrm{X}$ point, the magnon energy vanishes, accompanied by a divergent RIXS intensity~\cite{PhysRevLett.105.167404}. The RIXS spectral weight is zero at the $\Gamma$ point and becomes constant along the $\mathrm{M}-\mathrm{K}$ path. 

In Fig.~\ref{fig:fig4}(b), we investigate the influence of frustration parameters $(J_2,J_3)$ on the magnon dispersion and the RIXS intensity. As $(J_2,J_3)$ increases, the magnon energy is suppressed. Fig.~\ref{fig:fig4}(c) demonstrates the effect of the cyclic exchange interaction $J_c$, which primarily reduces the magnon energy near the $\mathrm{K}$ point. However, neither the frustration parameters $(J_2,J_3)$ nor the cyclic exchange $J_c$ significantly alters the single-magnon spectral weight.

\subsubsection{One-to-three magnon hybridization process}
Spin wave expansion is a perturbative method that enumerates quantum fluctuations which originate from magnon-magnon interaction. This leads to hybridization vertices arising from $H_1$ [Eq.~\eqref{eq:Svertex1}]. The vertices $V^{(4)}_{1234},V^{(5)}_{1234}$, and $V^{(6)}_{1234}$ between one- and three- magnon states contribute substantially to the renormalization of the excitation spectra of  the three-magnon continuum~\cite{PhysRevB.72.014403,PhysRevB.85.064421,PhysRevLett.115.207202,10.21468/SciPostPhys.4.1.001}. The three-magnon term $I^{(0,1/S,1)}_{\mathrm{3m}}$ originating from the $1/S$ expansion of the single spin-flip operator $S_i^x$ cannot be neglected, as it  will also contribute to the satellite intensity near the single-magnon  RIXS intensity~\cite{Igarashi_2015_effect,PhysRevB.85.064421,PhysRevB.48.3264}. In this section, we aim to investigate how the one-to-three magnon hybridization process contributes to the two-spin correlation function at $\mathcal{O}(\text{UCL}[0])$. In our case the four-spin cyclic interaction serves as a non-perturbative correction parameter~\cite{PhysRevB.79.235130}. The topologies of the Feynman diagrams belonging to $J_c$ and the hybridization vertices [Eq.~\eqref{eq:i1s3m}] are different from each other~\cite{PhysRevB.79.235130,canali1992theory}. In our calculations, we mainly focus on the RIXS spectrum arising from the one-to-three magnon hybridization process, and ignore contributions from the $J_c$ term.

The three-magnon DOS $D_{\mathrm{3m}}(\qq,\omega)$ [Eq.~\eqref{eq:d3m}] is the convolution of the three-magnon states~\cite{ament2010strong}, which is given by
\begin{align}
&D_{\mathrm{3m}}(\qq,\omega)=\frac{1}{N^2}\sum\limits_{\kk_1,\kk_2,\kk_3}
\delta(\omega-\omega_{\kk_1}-\omega_{\kk_2}-\omega_{\kk_3}),\label{eq:d3m}
\end{align}
where $\kk_1+\kk_2+\kk_3=\qq$. The three-magnon RIXS intensity expression $I^{(0,1/S,1)}_{\mathrm{3m}}(\qq,\omega)$ [Eq.~\eqref{eq:i1s3m}] at $\mathcal{O}(\text{UCL}[0])$ is given by  
\begin{align}
\notag
&I^{(0,1/S,1)}_{\mathrm{3m}}(\qq,\omega)
\propto \frac{1}{N^2S}(u_{\qq}-v_{\qq})^2\sum\limits_{\kk,\pp}\left|\mathrm{f^{(\mathsf{a})}_{3m}}(\kk,\pp,\qq)\right.\nonumber \\
&\left.
+\mathrm{f^{(\mathsf{b})}_{3m}}(\kk,\pp,\qq)+\mathcal{V}(\kk,\pp,\qq)\right|^2\delta(\omega-\omega_{\kk}-\omega_{\pp}-\omega_{\pp+\qq-\kk}),
\label{eq:i1s3m}
\end{align} calculated using the scattering operator Eq.~\eqref{eq:o03m}, with the vertex term~\cite{PhysRevB.72.014403,PhysRevB.85.064421}
\begin{align}
&\mathcal{V}(\kk,\pp,\qq)=\nonumber\\
&
2u_\qq u_\kk u_\pp  u_{\pp+\qq-\kk} \notag \\
&\left(\frac{V^{(4)}_{\qq,-\kk,-\pp,\pp+\qq-\kk}-\text{sign}(\gamma_\bg) V_{\qq,-\kk,-\pp,\pp+\qq-\kk}^{(6)}}{\omega_\kk+\omega_\pp+\omega_{\pp+\qq-\kk}}\right),
\end{align}
where \text{sign}$(\gamma_\bg)$ is a consequence of the umklapp process~\cite{PhysRevB.45.10131}. The scattering matrix elements $\mathrm{f^{(\mathsf{a})}_{3m}}(\kk,\pp,\qq)$ and $\mathrm{f^{(\mathsf{b})}_{3m}}(\kk,\pp,\qq)$ are stated in Appendix.~\ref{app:3msme}. 

In Fig.~\ref{fig:fig9}, we plot the three-magnon DOS and the three-magnon RIXS intensity from the $1/S$ expansion. In Fig.~\ref{fig:fig9}(a), the upper bound of the three-magnon DOS is $\omega \sim 3(2.3J_1)=6.9J_1$ with a flat feature. With the introduction of a finite amount of frustration, the upper bound of the energy is suppressed, see Fig.~\ref{fig:fig9}(b). In Fig.~\ref{fig:fig9}(c), we show the three-magnon RIXS intensity resulting from the $1/S$ expansion by incorporating the one-to-three magnon hybridization process. We note the presence of broad high-energy band features with slight intensity peaks near $3J_1$. The intensities around the $\mathrm{X}$ point vanish instead of diverging, which indicates that these intensities cannot be attributed to single-magnon dispersion, which are supposed to diverge at the $\mathrm{X}$ point (the ordering wave vector)~\cite{PhysRevLett.105.167404}. 

Next, we consider finite values of frustration  $(J_2,J_3)=0.05J_1$. The energy of the whole spectrum is downshifted, see Fig.~\ref{fig:fig9}(d). The intensity of the high-energy bands main peaks become more prominent. There is still a weak intensity peak above $\sim 2J_1$ in the $\mathrm{M}-\mathrm{K}$ path, which could potentially account for the satellite intensity peaks around the single-magnon dispersion peaks. However, based on our calculations, we conclude that the three-magnon intensity peaks in $\mathcal{O}(\text{UCL}[1])$ and $\mathcal{O}(\text{UCL}[2])$ (in Sec.~\ref{subsec:6scf}) do not display features of the satellite intensity peaks near $\sim 2J_1$. This is one of the reasons why we choose to analyze the model within a SMBFT formalism. Note, within SBMFT, the spin-flip operator $S^x_i$ is expressed as a bilinear combination of spinon operators~\cite{lacroix2011introduction}. Consequently, SBMFT is not inherently suited for computing the single spin-flip RIXS spectra, as it prioritizes to capture nonlocal and collective features over strictly local dynamics. However, this limitation does not preclude its use in evaluating operators that intrinsically reflect the emergent properties of spinons, such as the three-spin-flip scattering operator [Eq.~\eqref{eq:o4s}], which can reveal the spinon BEC spectrum (Fig.~\ref{fig:fig12}(c) in Sec.~\ref{subsubsec:4spi}). It can also explain the satellite intensity peaks of the single-magnon.

\begin{figure}[htbp]
\includegraphics[width=0.5\textwidth]{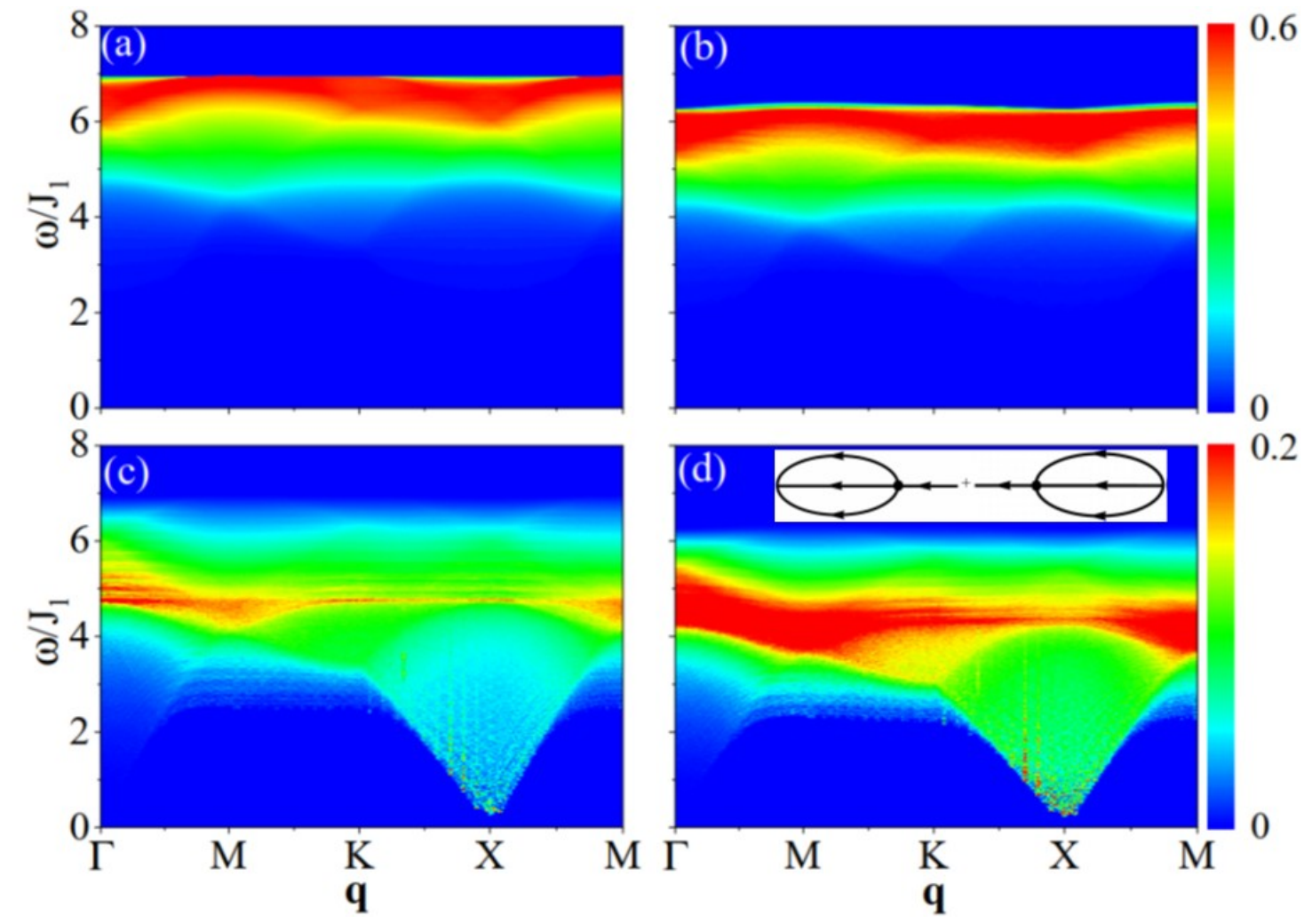} \\
\caption{Three-magnon DOS [Eq.~\eqref{eq:d3m}] in (a) and (b) and three-magnon intensity $I^{(0,1/S,1)}_{\mathrm{3m}}(\qq,\omega)$ [$\text{Eq.}\eqref{eq:i1s3m}$] in (c) and (d). Panels (a) and (c) use parameters $(J_2,J_3,J_c) = 0$, whereas panels (b) and (d) include frustration effects with $(J_2,J_3,J_c) = (0.05,0.05,0)J_1$. In (a) the upper bound of the three-magnon DOS is $\approx 6.9 J_1$. With finite frustration $(J_2,J_3) = (0.05,0.05)J_1$, this upper bound is lowered to $\approx 6.2 J_1$ in panel (b). Panel (c) presents the three-magnon intensity $I^{(0,1/S,1)}_{\mathrm{3m}}(\qq,\omega)$, where a broad high-energy band feature is evident. Upon incorporating the one-to-three hybridization process, the three-magnon excitation peaks shift toward the one-magnon excitation regions. Notably, the intensity peaks vanish at the $\mathrm{X}$ point instead of diverging, indicating that these peaks do not correspond to single-magnon dispersion peaks, which are known to have gapless dispersion and always diverge at the $\mathrm{X}$ point~\cite{PhysRevLett.105.167404}. With finite frustration, the overall spectrum is shifted to lower energies, and the high-energy band peaks become more pronounced. Additionally, some peaks remain near $\sim 2J_1$, suggesting the presence of satellite intensity peaks associated with single-magnon dispersion peaks. In panel (d), consistent with the DOS, the inclusion of frustrations causes lower energy excitations in RIXS spectrum. The inset in panel (d) is showing the Feynman diagram corresponding to the one-to-three magnon hybridization process.}
    \label{fig:fig9}
\end{figure}

\subsection{Four-spin correlation function\label{subsec:4scf}} 

We now proceed with the four-spin correlation RIXS spectrum calculation. First, we present the non-interacting two-magnon RIXS calculation, incorporating the UCL expansion corrections up to first order $I^{(1,S^2,0)}_{\mathrm{2m}}$ and second order $I^{(2,S^2,0)}_{\mathrm{2m}}$where $\mathcal{L}=1,2$ and $\mathcal{N}=0$ in Eq.~\eqref{eq:stgreen}. We used scattering operators in Eqs.~\eqref{eq:o1s2m} and \eqref{eq:2ucl4s} to derive the intensity expressions. Next, we examine the bimagnon RIXS spectrum within the ladder approximation. Finally, we present and discuss the four-spin correlation function spectrum calculated using SBMFT and compare the results with those from the local spin-flip schemes.

\subsubsection{Two-magnon RIXS spectrum }\label{subsubsec:2mrs}
Spectroscopic behavior of the four-spin correlation function, which manifests in the two-magnon RIXS intensity, has been studied both at the $K$-~\cite{van2007theory,nagao2007two,PhysRevB.77.134428,luo2014spectrum} and the $L$-edge ~\cite{PhysRevB.85.064421,pal2023theoretical}. Creating the two-magnon continuum involves generating a two spin-flip processes. To provide a comprehensive understanding of the two-magnon spectrum, we compute the two-magnon DOS $D_{\mathrm{2m}}(\qq,\omega)$ given by the expression 
\begin{align}
\label{eq:d2m}
D_{\mathrm{2m}}(\qq,\omega)=\frac{1}{N}\sum\limits_\kk \delta(\omega-\omega_{\kk+\frac{\qq}{2}}-\omega_{\kk-\frac{\qq}{2}}).
\end{align} 
Combining Eq.~\eqref{eq:o1s2m} with Eq.~\eqref{eq:stgreen}, we compute the two-magnon RIXS scattering response at $\mathcal{O}(\text{UCL}[1])$ to obtain
\begin{align}
\label{eq:i12m}
I^{(1,S^2,0)}_{\mathrm{2m}}(\qq,\omega)=\frac{S^2}{N}\sum\limits_\kk [\mathrm{f^{(1)}_{\mathrm{2m}}}(\kk,\qq)]^2\delta(\omega-\omega_{\kk+\frac{\qq}{2}}-\omega_{\kk-\frac{\qq}{2}}).
\end{align}
\begin{figure}[htb]
\centering
\includegraphics[width=\linewidth,scale=1.00]{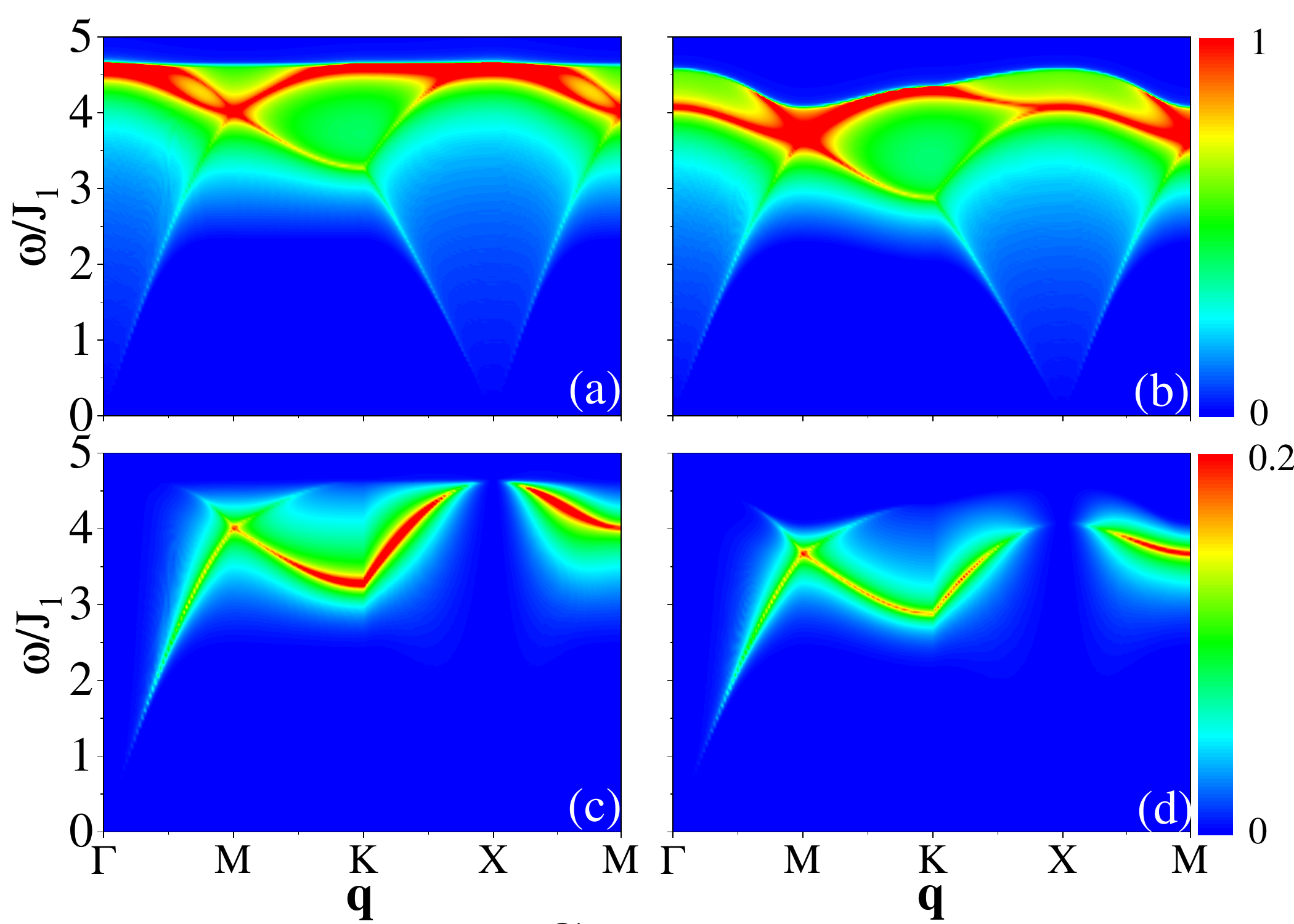}
\caption{Two-magnon DOS $D_{\mathrm{2m}}$ [Eq.~\eqref{eq:d2m}] (the upper panels) and RIXS intensity $I^{(1,S^2,0)}_{\mathrm{2m}}$ [Eq.~\eqref{eq:i12m}] (the bottom panels) within $\mathcal{O}(\text{UCL}[1])$. (a) Two-magnon DOS with $(J_2,J_3,J_c)=0$ shows a flat upper bound at $\approx 4.6J_1$. (b) Two-magnon DOS with $(J_2,J_3,J_c)=(0.05,0.05,0.4) J_1$. The upper bound becomes wavy with a local minimum at the $\mathrm{M}$ point. (c) Two-magnon RIXS intensity with $(J_2,J_3,J_c)=0$. (d) Two-magnon RIXS intensity with $(J_2,J_3,J_c)=(0.05,0.05,0.4)J_1$. Two-magnon RIXS intensity vanishes at $\mathrm{\Gamma}$ and $\mathrm{X}$ points in $\mathcal{O}(\text{UCL}[1])$~\cite{PhysRevB.77.134428}. The intensity peak at $\mathrm{M}$ point has higher energy than that at $\mathrm{K}$ point. Moving to $\mathrm{X}$ point, the energy of two-magnon intensity becomes $\approx 4.6J_1$ instead of $0$, which indicates the two-magnon intensity is not diverging at the $\mathrm{X}$ point. Comparing (c) and (d), the inclusion of $(J_2,J_3,J_c)$ leads to a downward shift in the energy range and a softening of the peaks. } 
 \label{fig:fig5}
\end{figure}

In Fig.~\ref{fig:fig5}, we show the results of $D_{\mathrm{2m}}(\qq,\omega)$ [Eq.~\eqref{eq:d2m}] and $I^{(1,S^2,0)}_{\mathrm{2m}}(\qq,\omega)$ [Eq.~\eqref{eq:i12m}]. For the model with only $J_1$, the two-magnon DOS shows a flat upper bound at $\approx 4.6 J_1$,  around twice the energy of the flat band in the single-magnon dispersion, see Fig.~\ref{fig:fig5}(a). With the inclusion of frustration $(J_2,J_3)$ and cyclic exchange interactions $J_c$, the two-magnon DOS shifts toward lower energies (Fig.~\ref{fig:fig5}(b)), consistent with the lower magnon energy in Figs.~\ref{fig:fig4}(b)-(c). In Fig.~\ref{fig:fig5}(b), the upper bound becomes undulated with a local minimum at the $\mathrm{M}$ point, which has the greatest concentration of energy states. The corresponding two-magnon RIXS spectra are presented in Figs.~\ref{fig:fig5}(c)-\ref{fig:fig5}(d). We observe that the two-magnon RIXS intensity vanishes at the $\mathrm{\Gamma}$ and the $\mathrm{X}$ points within $\mathcal{O}(\text{UCL}[1])$~\cite{PhysRevB.77.134428}. In such a spin-conserving process, the spin-spin correlations do not contribute to the $\mathrm{X}$ point, leading to zero intensity. The intensity peak around $\mathrm{M}$ has a higher energy than at $\mathrm{K}$. 

Next, we discuss the RIXS intensity $I^{(2,S^2,0)}_{\mathrm{2m}}(\qq,\omega)$ at $\mathcal{O}(\text{UCL}[2])$. There are three spectroscopic contributions at this level. They are given by $I^{(2,S^2,0)}_{\mathrm{2m}}(\qq,\omega)=c_1I^{(2_{ij},S^2,0)}_{\mathrm{2m}}(\qq,\omega)+c_2 I^{(2_{jl},S^2,0)}_{\mathrm{2m}}(\qq,\omega) 
    +c_3I^{([2_{ij},2_{jl}],S^2,0)}_{\mathrm{2m}}(\qq,\omega)$, where $i,j$, and $l$ refer to lattice sites.  Here $(c_1,c_2,c_3) = (\frac{1}{4},\frac{1}{16},\frac{1}{4})$ are identified from Eq.~\eqref{eq:4s} in Appendix~\ref{app:msme}. In the first term, $I^{(2_{ij},S^2,0)}_{\mathrm{2m}}(\qq,\omega)$ is mathematically the same expression as Eq.~\eqref{eq:i12m}, the intensity at $\mathcal{O}(\text{UCL}[1])$. The second term 
\begin{align}
 I^{(2_{jl},S^2,0)}_{\mathrm{2m}}(\qq,\omega)=\frac{S^2}{N}\sum\limits_\kk [\mathrm{f^{(2_{jl})}_{2m}}(\kk,\qq)]^2\delta(\omega-\omega_{\kk}-\omega_{\kk-\qq}),
 \label{eq:i22m}
 \end{align}
 is derived using the scattering operator Eq.~\eqref{eq:2ucl4s}. The third term 
\begin{align}
      \notag 
      &I^{([2_{ij},2_{jl}],S^2,0)}_{\mathrm{2m}}(\qq,\omega)=\frac{S^2}{N}\sum\limits_\kk [\mathrm{f^{(2_{ij})}_{2m}}(\kk,\qq)\mathrm{f^{(2_{jl})}_{2m}}(\kk,\qq)] \nonumber \\
      &\left[\delta(\omega-\omega_{\kk+\frac{\qq}{2}}-\omega_{\kk-\frac{\qq}{2}})+\delta(\omega-\omega_{\kk}-\omega_{\kk-\qq})\right],
      \label{eq:i122m}
\end{align} is the cross-term contribution from the previous two. This is based on the form of the scattering operator given in Eq.~\eqref{eq:4s}. The derivation of the RIXS intensity formula is outlined in Appendix.~\ref{app:ri}. Finally, the total two-magnon RIXS intensity up to $\mathcal{O}(\text{UCL}[2])$ is given by
\begin{align}
I^{\text{tot}}_{\mathrm{2m}}(\qq,\omega)&=  \xi^2I^{(1,S^2,0)}_{\mathrm{2m}}(\qq,\omega)+ \xi^4I^{(2,S^2,0)}_{\mathrm{2m}}(\qq,\omega).
\label{eq:itot2m}
\end{align}

\begin{figure}[htbp]
\centering
\includegraphics[width=85mm]{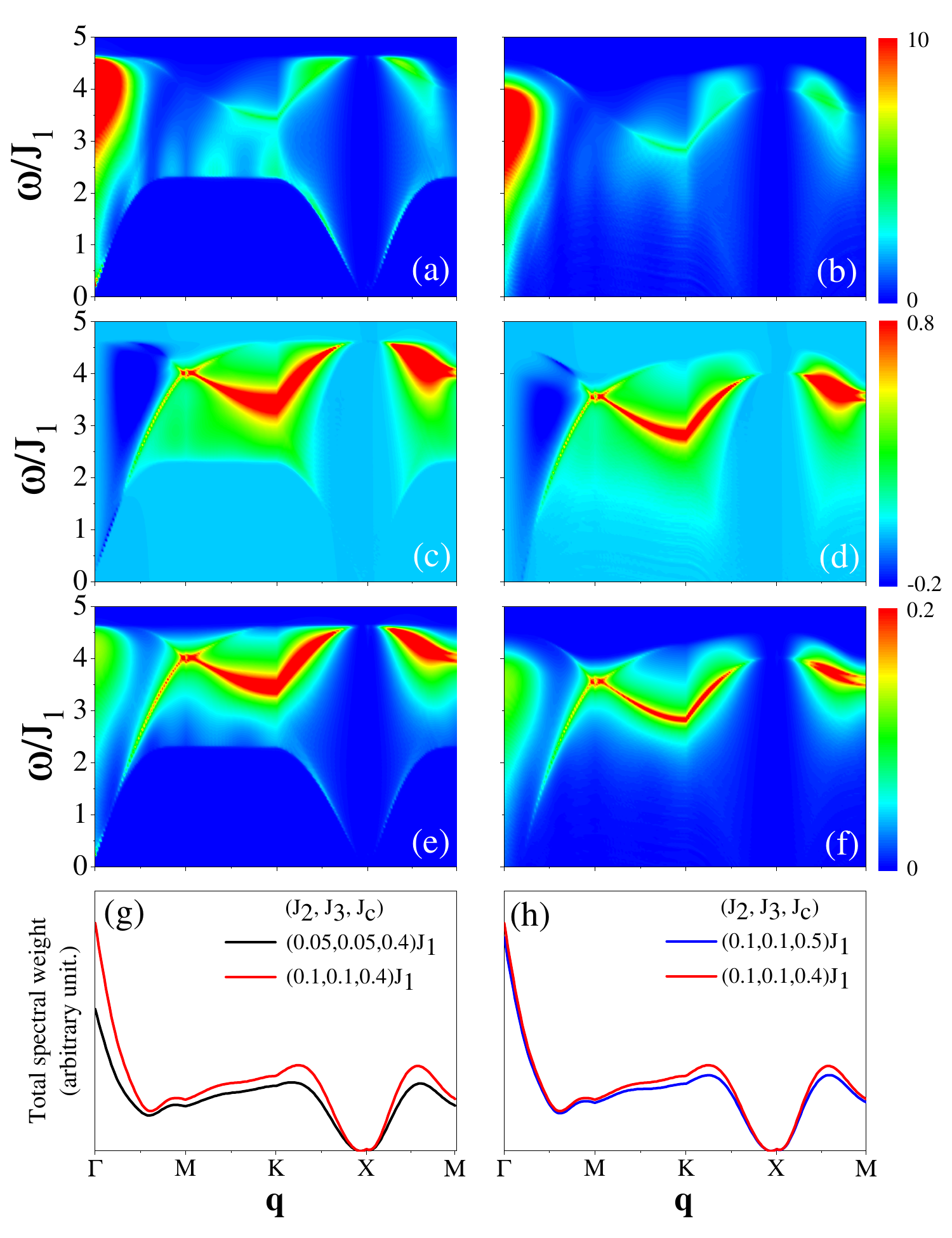}
\caption{Two-magnon RIXS intensity within $\mathcal{O}(\text{UCL}[2])$. Panels (a),(c),and (e) correspond to the case with $(J_2,J_3,J_c) = 0$, while panels (b), (d), and (f) depict results with $(J_2,J_3,J_c) = (0.05,0.05,0.4)J_1$. (a) and (b) Two-magnon intensity $I^{(2_{jl},S^2,0)}_{\mathrm{2m}}(\qq,\omega)$ [Eq.~\eqref{eq:i22m}]. In both cases, large RIXS intensities appear around the $\qq=\Gamma$ point. For $(J_2,J_3,J_c) = 0$, the low-energy intensity outlines the single-magnon dispersion and vanishes at the $\mathrm{X}$ point. However, for finite $(J_2,J_3,J_c)$ in panel (b), the single-magnon dispersion feature disappears, and the entire spectrum shifts to lower energies. (c) and (d) Two-magnon intensity $I^{([2_{ij},2_{jl}],S^2,0)}_{\mathrm{2m}}(\qq,\omega)$ [Eq.~\eqref{eq:i122m}]. Near the $\Gamma$ point, negative intensity regions appear.  (e) and (f) The total RIXS intensity $I^{\text{tot}}_{\mathrm{2m}}(\qq,\omega)$ [Eq.~\eqref{eq:itot2m}]. The intensity peaks at $\qq=\Gamma$ point come from $I^{(2_{jl},S^2,0)}_{\mathrm{2m}}(\qq,\omega)$. The total RIXS spectrum is not continuous due to the negative contributions from $I^{([2_{ij},2_{jl}],S^2,0)}_{\mathrm{2m}}(\qq,\omega)$ . (g) and (h) Two-magnon RIXS spectral weight $W^{\text{tot}}_{\mathrm{2m}}(\qq)$. The parameters are $(J_2,J_3,J_c)=(0.05,0.05,0.4)J_1$, $(0.1,0.1,0.4)J_1$, and $(0.1,0.1,0.5)J_1$ in black,red,blue curve, respectively. The big weight at $\qq=\Gamma$ point refer to summation weights of peaks with high energy and nearly zero energy.} 
 \label{fig:fig6}
\end{figure}

Since the first term in $I^{(2,S^2,0)}_{\mathrm{2m}}(\qq,\omega)$ is proportional to $\mathcal{O}(\text{UCL}[1])$, which has been presented in Fig.~\ref{fig:fig5}, we show the remaining contributions of two-magnon RIXS intensity up to $\mathcal{O}(\text{UCL}[2])$ in Fig.~\ref{fig:fig6}. Figs.~\ref{fig:fig6}(a) and \ref{fig:fig6}(b) show the contribution of the $I^{(2_{jl},S^2,0)}_{\mathrm{2m}}(\qq,\omega)$ term [Eq.~\eqref{eq:i22m}]. We observe a substantial large intensity at the $\mathrm{\Gamma}$ point which arises from spin correlation between nearest neighbors $\bs_j\cdot \bs_l$ (since this term does not commute with the Hamiltonian at $\qq=0$)~\cite{PhysRevB.77.134428,pal2023theoretical}. For the model with only $J_1$, see Fig.~\ref{fig:fig6}(a), there are two spectrum branches in which the low-energy branch outlines the single-magnon dispersion and vanishes at $\qq=\mathrm{X}$. Inclusion of finite $(J_2,J_3,J_c)$ introduces a downshift in energy which leads to the disappearance of the low-energy branch as seen in Fig.~\ref{fig:fig6}(b). Additionally, Figs.~\ref{fig:fig6}(c) and \ref{fig:fig6}(d) display the contribution of the cross term $I^{([2_{ij},2_{jl}],S^2,0)}_{\mathrm{2m}}(\qq,\omega)$ [Eq.~\eqref{eq:i122m}]. For this term (which is a part of the total spectrum), a minor negative intensity contribution is obtained near the $\mathrm{\Gamma}$ point (see Eq.~\ref{eq:i122m}). The positive parts reproduce most of the high-energy features of the two-magnon spectrum within $\mathcal{O}(\text{UCL}[1])$ (see Figs.~\ref{fig:fig5}(c) and (d)). The inclusion of finite $(J_2,J_3,J_c)$ also leads to a downshift of energy and the disappearance of the low-energy branch, see Fig.~\ref{fig:fig6}(d).  

In Figs.~\ref{fig:fig6}(e)-\ref{fig:fig6}(f) we give the \emph{total} RIXS spectra $I^{\text{tot}}_{\mathrm{2m}}(\qq,\omega)$ [Eq.~\eqref{eq:itot2m}] up to $\mathcal{O}(\text{UCL}[2])$ which is overall positive. In the total spectrum, the nonzero intensity near the $\Gamma$ point is contributed by the $I^{(2_{jl},S^2,0)}_{\mathrm{2m}}(\qq,\omega)$ term, while the discontinuous spectrum arising in the $\Gamma-\mathrm{M}$ momentum path is due to the negative intensity contribution from the $I^{([2_{ij},2_{jl}],S^2,0)}_{\mathrm{2m}}(\qq,\omega)$ term.  In Figs.~\ref{fig:fig6}(g)-\ref{fig:fig6}(h) we plot the total two-magnon spectral weight $W^{\text{tot}}_{\mathrm{2m}}(\qq)$ up to $\mathcal{O}(\text{UCL}[2])$ using Eq.~\eqref{eq:weight}. 
The $I^{\text{tot}}_{\mathrm{2m}}(\qq,\omega)$ response has a dominant intensity at the $\qq=\Gamma$ point, including peaks with $\omega\sim 0$. At the $\qq=\mathrm{X}$ point, the spectral weight becomes zero since there is a zero two-magnon intensity.

The negative RIXS intensity peaks originating from the cross-term contribution in $\mathcal{O}(\text{UCL}[2])$ could potentially be cured by including higher order UCL or spin-wave corrections. However, such an approach is neither practical nor analytically tractable as evident from the complications of the calculation at the current level. Later in Sec.~\ref{subsubsec:2spi}, we will demonstrate, that it is possible to reproduce the four-spin correlation spectrum up to $\mathcal{O}(\text{UCL}[2])$ by computing the two spin-flip RIXS spectrum within the SBMFT context at $\mathcal{O}(\text{UCL}[1])$ itself. We will demonstrate that not only is the calculation of RIXS from this viewpoint more manageable, additionally it offers a deeper physical insight into the effects of higher order spin correlations and entanglement effects arising from nearest-neighbor lattice points via a fluctuating RVB bond spin-flip RIXS scheme.

\subsubsection{Bimagnon RIXS spectrum}\label{subsubsec:bi}
Bimagnon is a coherent bound state of the two-magnon continuum. The magnon-magnon interactions create local spin rearrangements that weaken the magnetic RIXS spectrum peaks at large $\qq$ scales, but have minimal impact at $\qq \sim 0$~\cite{PhysRevB.77.134428,PhysRevLett.102.167401,PhysRevB.75.020403}. In the two-magnon RIXS case, the influence of the magnon-magnon interaction can be calculated within the Bethe-Salpeter approach after summing the ladder diagram series~\cite{luo2014spectrum,xiong2017magnon,PhysRevB.100.054410,nagao2007two}. The vertex terms can be expanded into the interaction potential with respect to $\qq$ in 18 channels (10 channels if $(J_2,J_3,J_c)=0$) of Bogoliubov quasiparticles as
\begin{equation}
\frac{1}{S}u_1u_2u_3u_4V^{(3)}_{1234}=\hat{v}(3,2)\hat{\Gamma}(\qq)\hat{v}^T(1,4).
\end{equation} 
The general expression for the bimagnon scattering channels $\hat{v}$ and the interacting potentials $\hat{\Gamma}(\qq)$ have been presented in our former papers~\cite{luo2014spectrum,xiong2017magnon}. Using those expressions and considering the Dyson equation, after summing the ladder diagrams exactly, we have the two-magnon Green's function $G(\qq,\omega)$ given by
\begin{align}
G(\qq,\omega)\notag&=G_0(\qq,\omega)\\
&+\hat{g}(\qq,\omega)\hat{\Gamma}(\qq)[I-\hat{R}(\qq,\omega)\hat{\Gamma}(\qq)]^{-1}\hat{g}^T(\qq,\omega),
\end{align}
where
\begin{subequations}
\begin{eqnarray}
\hat{g}^{(i)}(\qq,\omega)&=&\frac{2}{N}\sum\limits_\kk f^{(i)}_{\mathrm{2m}}\Pi_{\mathrm{2m}}^{(i)}(\qq,\omega)\hat{v}^{(i)},\\
\hat{R}^{(i)}(\qq,\omega)&=&\frac{2}{N}\sum\limits_\kk\left(\hat{v^{(i)}}\right)^T\Pi^{(i)}_{\mathrm{2m}}(\qq,\omega)\hat{v}^{(i)},
\end{eqnarray}
\end{subequations}
with $i=1,2_{jl}$. The two-magnon propagators are 
\begin{subequations}
\begin{eqnarray}
\Pi_{\mathrm{2m}}^{(1)}(\qq,\omega)&=&\left(\omega-\omega_{\kk+\frac{\qq}{2}}-\omega_{\kk-\frac{\qq}{2}}+i0^+\right)^{-1},\\
\Pi_{\mathrm{2m}}^{(2_{jl})}(\qq,\omega)&=&\left(\omega-\omega_{\kk}-\omega_{\kk-\qq}+i0^+\right)^{-1}.
\end{eqnarray}
\end{subequations}
Using Eq.~\eqref{eq:stgreen}, the magnon-magnon interaction contribution to the bimagnon RIXS intensity is expressed as
\begin{align}
&I^{(i,S^2,0),\text{int}}_{2m}(\qq,\omega)\notag\\
&=-\frac{1}{\pi}\mathrm{Im}\left[\hat{g}^{(i)}(\qq,\omega)\frac{\hat{\Gamma}(\qq)}{I-\hat{R}^{(i)}(\qq,\omega)\hat{\Gamma}(\qq)]}\left(\hat{g}^{(i)}(\qq,\omega)\right)^T\right].
\end{align}
In the above, the superscript ``int" denotes interaction. The resulting bimagnon RIXS intensity can be expressed as
\begin{align}
I^{(i,S^2,0)}_{\mathrm{bm}}(\qq,\omega)=I^{(i,S^2,0)}_{\mathrm{2m}}(\qq,\omega)+I^{(i,S^2,0),\text{int}}_{\mathrm{2m}}(\qq,\omega),
\end{align} 
where the symbol bm denotes bimagnon. $i=1, 2_{ij}, 2_{jl}$ represents the order of UCL expansion. Note, that the $2_{ij}$ term has the same mathematical expression as the $i=1$ term.

\begin{figure}[t]
\centering
\includegraphics[width=3.4in]{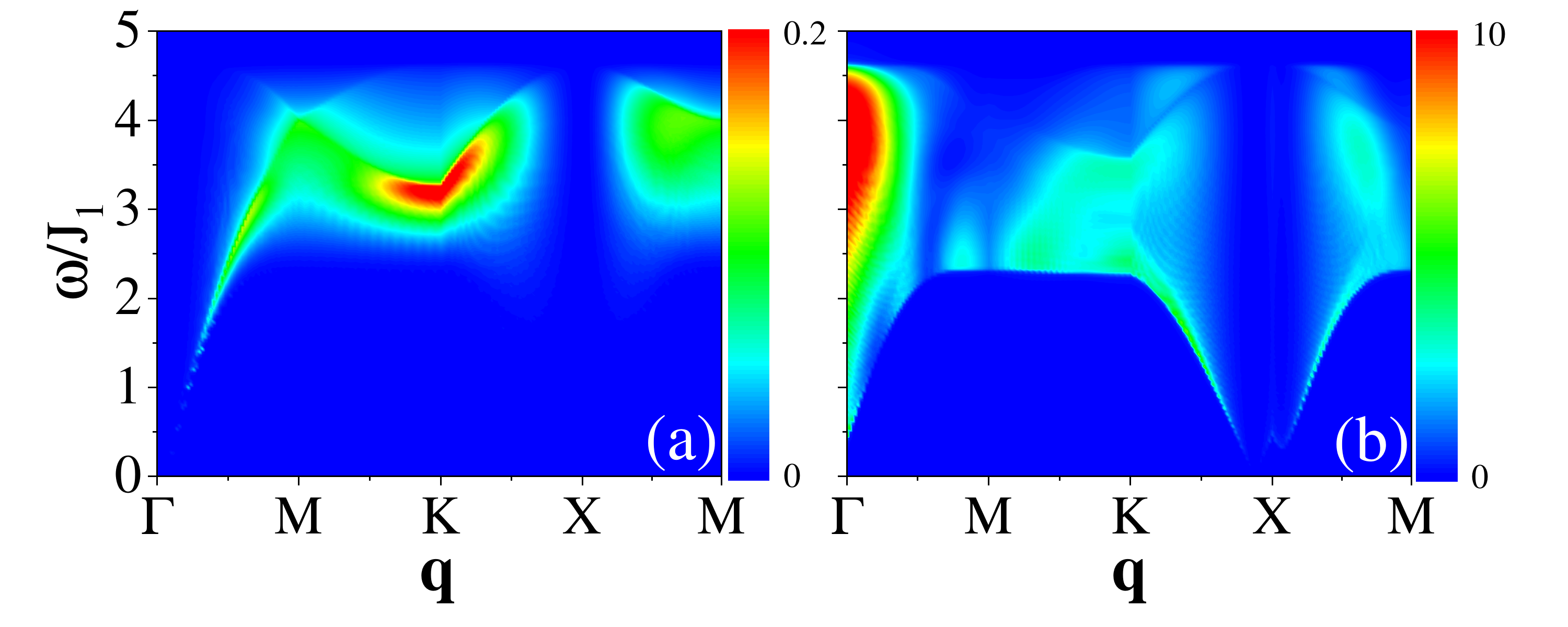}
\caption{Bimagnon RIXS spectrum. Magnon-magnon interaction effects are incorporated within the ladder approximation scheme. Up to second order there are two contributions to bimagnon RIXS intensity. (a) $I^{(1,S^2,0)}_{\mathrm{bm}}$ is constructed out of the two-magnon RIXS intensity $I^{(1,S^2,0)}_{\mathrm{2m}}$ [Eq.~\eqref{eq:i12m}] in $\mathcal{O}(\text{UCL}[1])$. (b) $I^{(2_{jl},S^2,0)}_{\mathrm{bm}}$ is computed out of the two-magnon RIXS intensity $I^{(2_{jl},S^2,0)}_{\mathrm{2m}}$ [Eq.~\eqref{eq:i22m}] in $\mathcal{O}(\text{UCL}[1])$ The magnon-magnon interaction corrections result in further softened peaks in general, with minimal impact near the $\qq=\Gamma$ point. The parameters are $(J_2,J_3,J_c)=0$. }
\label{fig:fig7}
\end{figure}

In Fig.~\ref{fig:fig7} we present the bimagnon RIXS spectrum building on the two-magnon states within $\mathcal{O}(\text{UCL}[2])$. Figs.~\ref{fig:fig7}(a) and \ref{fig:fig7}(b) are the bimagnon RIXS spectrum $I^{(1,S^2,0)}_{\mathrm{bm}}$ and $I^{(2_{jl},S^2,0)}_{\mathrm{bm}}$  incorporating the magnon-magnon interaction effects with self-energy correction from the Dyson equations, respectively. Comparing Fig.~\ref{fig:fig7} with Figs.~\ref{fig:fig6}(a) and \ref{fig:fig6}(c), the sharp peaks soften and broaden along the momentum path $\mathrm{M}-\mathrm{X}$ when magnon-magnon interaction is included. Some intensities show small red-shifts, consistent with previous theoretical calculations~\cite{nagao2007two,luo2014spectrum}. Due to the local magnetic screening effects and higher order correction of the UCL expansion~\cite{PhysRevLett.102.167401}, the intensity near the $\mathrm{\Gamma}$ point becomes more visible. The magnon-magnon interaction do not cause any red-shift to the $\qq\sim 0$ spectrum. The coherency of the two-magnon continuum clearly manifests at large $\qq$. We note the softened RIXS spectra is due to the vertex-corrected bimagnon and not the higher order spin excitations. Note, when using ladder diagram methods to compute magnon-magnon interaction effects in RIXS, the vertices arising from the cross-term in Eq.~\eqref{eq:ct}, have quasi-particle momenta that is not conserved during the creation and annihilation process. We ignore such momentum non-conserving terms in our ladder diagram calculation. 

\subsubsection{SBMFT results of mean-field two-spinon response\label{subsubsec:2spi}}
The SBMFT formalism intrinsically includes magnon-magnon interaction effects since it handles four-boson terms at the mean-field level~\cite{pires2021theoretical}. In the RIXS SC channel, the mean-field two-spinon spectrum consists of two parts -- a singular part and a continuum part. The singular part is given by the BEC of the spinon which results in a state with broken $SU(2)$ symmetry, yielding the quantum-corrected N\'eel ordering. The continuum part is given by the mean-field two-spinon continuum preserving $SU(2)$ symmetry with enhanced quantum fluctuation effects. The RIXS intensity expressions of the continuum $I^c_{\mathrm{2s}}(\qq,\omega)$ and the singular $I^s_{\mathrm{2s}}(\qq,\omega)$ parts are given by 
\begin{subequations}
\begin{eqnarray}
\notag  I^c_{\mathrm{2s}}(\qq,\omega)\propto \frac{1}{N}\sum\limits_{\substack{\kk\notin \pm \frac{\bq}{2},\\ \kk\in BZ}}|\mathrm{f_{2sc}}(\kk,\qq)|^2\delta(\omega-\omega^s_{\kk+\frac{\qq}{2}}-\omega^s_{\kk-\frac{\qq}{2}}),\\
\label{eq:ic2s}
\end{eqnarray}
\begin{eqnarray}
\notag &I^s_{\mathrm{2s}}(\qq,\omega)\propto N m_0\left\{|\mathrm{f_{2ssp}}(\qq)|^2\delta(\omega-\omega^s_{\frac{\bq}{2}+\frac{\qq}{2}}-\omega^s_{\frac{\bq}{2}-\frac{\qq}{2}})\right.\\
 &\left.+|\mathrm{f_{2ssm}}(\qq)|^2\delta(\omega-\omega^s_{-\frac{\bq}{2}+\frac{\qq}{2}}-\omega^s_{-\frac{\bq}{2}-\frac{\qq}{2}})\right\},
\label{eq:is2s}
\end{eqnarray}
\end{subequations}
where the scattering operator is listed in Eq.~\eqref{eq:o2smf}. In the above $\mathrm{f_{2sc}}(\kk,\qq)$ is the scattering matrix element of the continuum part. Here $\mathrm{2s}$ stands for two-spinon and $\mathrm{c}$ for continuum. The scattering matrix elements of the singular part are given by $\mathrm{f_{2ssp}(\qq)}$ and $\mathrm{f_{2ssm}}(\qq)$, where $\mathrm{2ssp} 
(\mathrm{2ssm})$ denotes the two-spinon singular part plus and minus contributions, respectively. The derivation details are outlined in Appendix~\ref{app:ssme}. The RIXS four-spin correlation function spectrum can be described by the response of the mean-field two-spinon continuum. This connection emphasizes the role of RVB fluctuations in the four-spin correlation spectrum that exist in the quantum corrected N\'{e}el state, which is protected by gauge symmetry. In this state, the massless gauge fluctuations confine spinons with enhanced quantum fluctuations beyond mean-field~\cite{kim1999theory,PhysRevB.52.440,jacobsen2010exact}.

\begin{figure}[htbp]
\includegraphics[width=85mm]{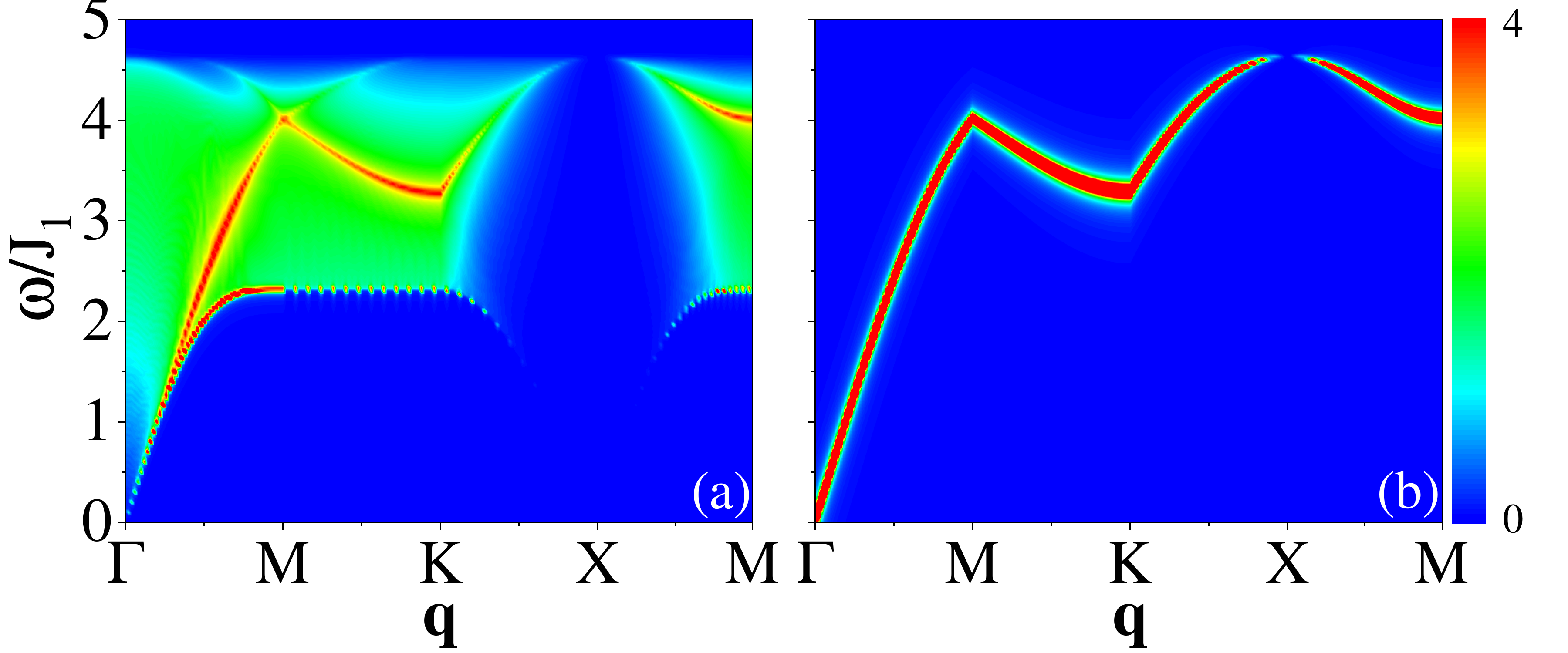}
\caption{Mean-field two-spinon $L_3$-edge RIXS spectrum  at zero temperature with $(J_2,J_3,J_c)=0$. (a) Continuum part $I^c_{\mathrm{2s}}(\qq,\omega)$ [Eq.~\eqref{eq:ic2s}]. (b) Singular part $I^s_{\mathrm{2s}}(\qq,\omega)$ [Eq.~\eqref{eq:is2s}]. The continuum part spectrum $I^c_{\mathrm{2s}}(\qq,\omega)$ is similar to the two-magnon RIXS spectrum in $\mathcal{O}(\text{UCL}[2])$ as displayed in Fig.~\ref{fig:fig6}(e). The SBMFT plot shows more continuous spectrum along the $\Gamma-\mathrm{M}$ path, indicating that spin-spin correlations described by higher order UCL expansion can be captured by RVB fluctuations (see Fig.~\ref{fig:fig2}(b)). The singular part of the spectrum $I^s_{\mathrm{2s}}(\qq,\omega)$ appears as a sharp peak in the continuum part. The mean-field two-spinon condensate response manifests as sharp peaks in the four-spin correlation spectra. The spectrum vanishes at ($\pi,\pi$), indicating the absence of a single-particle dispersion.}
\label{fig:fig8}
\end{figure}

In Fig.~\ref{fig:fig8}(a), we show the continuum part of the mean-field two-spinon spectrum (in the SC channel). Compared with the two-magnon spectrum up to $\mathcal{O}(\text{UCL}[2])$ shown in Fig.~\ref{fig:fig6}(e), it is broader and continuous. A non-zero intensity appears near the single-magnon dispersion which vanishes at the $\mathrm{X}$ point. An obvious intensity line is visible at high energies. 
Therefore, the mean-field two-spinon RIXS spectrum up to $\mathcal{O}(\text{UCL}[1])$ is able to capture all features of the two-magnon spectrum up to $\mathcal{O}(\text{UCL}[2])$. This suggests that the spin-spin correlation physics described by higher order UCL expansion (see scattering operator Eq.~\eqref{eq:2ucl4s}) can be detected by incorporating fluctuations of the RVB state. The bond spin-flip process inherently encompasses the three-site spin correlation as conceptualized in Fig.~\ref{fig:fig2}(b). In Fig.~\ref{fig:fig8}(b) we show the results of the singular part of the mean-field two-spinon RIXS spectrum.  A sharp and narrow intensity curve is present, similar to that shown in Fig.~\ref{fig:fig8}(a) and Fig.~\ref{fig:fig6}(e). This can be understood by the fact that the BEC of the mean-field two-spinon continuum breaks the $SU(2)$ symmetry, generating the quantum-corrected N\'{e}el state. 

\subsection{Six-spin correlation function \label{subsec:6scf}}

In the N\'eel antiferromagnetic state, the six-spin correlation function can contribute to the three-magnon RIXS response within the local spin-flip RIXS mechanism and to the four-spinon RIXS spectrum in the bond spin-flip mechanism. In the bond-spin-flip RIXS mechanism, the six-spin correlation function can be approximated by a mean-field treatment, leading to a four-spinon RIXS response. Furthermore, in this section, we compare the RIXS spectra of the three-magnon excitation with the mean-field four-spinon approximation. 

\subsubsection{Three-magnon continuum\label{subsubsec:3mag}}
The $\mathcal{O}(\text{UCL}[0])$ contribution to the three-magnon spectrum includes only $S^x$ in the scattering operator [Eq.~\eqref{eq:oql}]. This has been computed using Eq.~\eqref{eq:i1s3m} and is shown in Fig.~\ref{fig:fig9}. The $\mathcal{O}(\text{UCL}[1])$ and $\mathcal{O}(\text{UCL}[2])$ terms contribute up to the $S^3$ order in the scattering operator. These higher-order terms give rise to six-spin correlation functions in the RIXS intensity. They have been investigated and found to produce continuous intensity at high energies~\cite{ament2010strong,pal2023theoretical}. For the six-spin correlation function, the three-magnon $\mathcal{O}(\text{UCL}[1])$ intensity is given by
\begin{align}
&I^{(1,S^3,1)}_{\mathrm{3m}}(\qq,\omega)=  \frac{S^3}{N^2}\sum\limits_{\kk,\pp}\left\{[\mathrm{f^{(\mathsf{c})}_{3m}}(\kk,\pp,\qq)]^2+[\mathrm{f^{(\mathsf{d})}_{3m}}(\kk,\pp,\qq)]^2\right\} \nonumber \\
&\delta(\omega-\omega_{\kk+\pp+\qq}-\omega_{\pp}-\omega_{\kk}),
\label{eq:i33m}
\end{align}
 based on the RIXS scattering operator Eq.~\eqref{eq:o13m}.
In $\mathcal{O}(\text{UCL}[2])$, similar to the two-magnon RIXS intensity in Sec.~\ref{subsubsec:2mrs}, the three-magnon RIXS intensity is expressed as a sum of three terms: $I^{(2,S^3,1)}_{\qq,\mathrm{3m}}(\qq,\omega)=c_1I^{(2_{ij},S^3,1)}_{\mathrm{3m}}(\qq,\omega)+c_2 I^{(2_{jl},S^3,1)}_{\mathrm{3m}}(\qq,\omega) 
+c_3I^{([2_{ij},2_{jl}],S^3,1)}_{\mathrm{3m}}(\qq,\omega)$, where $(c_1,c_2,c_3) = (\frac{1}{4},\frac{1}{16},\frac{1}{4})$ are identified from Eq.~\eqref{eq:4s}. The second term, derived using the scattering operator Eq.~\eqref{eq:o23m} is
\begin{align}
\notag
&I^{(2_{jl},S^3,1)}_{\mathrm{3m}}(\qq,\omega)= \frac{S^3}{N^2}\sum\limits_{\kk,\pp}\left\{[\mathrm{f^{(\mathsf{e})}_{3m}}(\kk,\pp,\qq)]^2+[\mathrm{f^{(\mathsf{f})}_{3m}}(\kk,\pp,\qq)]^2\right\}\nonumber \\
&\delta(\omega-\omega_{\kk+\qq-\pp}-\omega_{\pp}-\omega_{\kk}). 
\label{eq:i43m}
\end{align}
The third term represents the cross-term contribution from the previous two and is given by
\begin{align}
\notag
&I^{([2_{ij},2_{jl}],S^3,1)}_{\mathrm{3m}}(\qq,\omega)=\\
& \frac{S^3}{N^2}\sum\limits_{\kk,\pp}\left\{\mathrm{f^{(\mathsf{c})}_{3m}(\kk,\pp,\qq)f^{(\mathsf{e})}_{3m}(\kk,\pp,\qq)+f^{(\mathsf{d})}_{3m}(\kk,\pp,\qq)f^{(\mathsf{f})}_{3m}}(\kk,\pp,\qq)\right\}\nonumber \\
&\left\{\delta(\omega-\omega_{\kk+\pp+\qq}-\omega_{\pp}-\omega_{\kk})+\delta(\omega-\omega_{\kk+\qq-\pp}-\omega_{\pp}-\omega_{\kk})\right\}.
\label{eq:i343m}
\end{align}
Thus, the total three-magnon RIXS intensity from the six-spin correlation is 
\begin{align}
I^{S^3,\text{tot}}_{\mathrm{3m}}(\qq,\omega)&=\xi^2 I^{(1,S^3,1)}_{\mathrm{3m}}(\qq,\omega)+ \xi^4 I^{(2,S^3,1)}_{\mathrm{3m}}(\qq,\omega). 
\label{eq:itot3m}
\end{align}

\begin{figure}[htbp]
\centering
\includegraphics[width=\linewidth,scale=1.00]{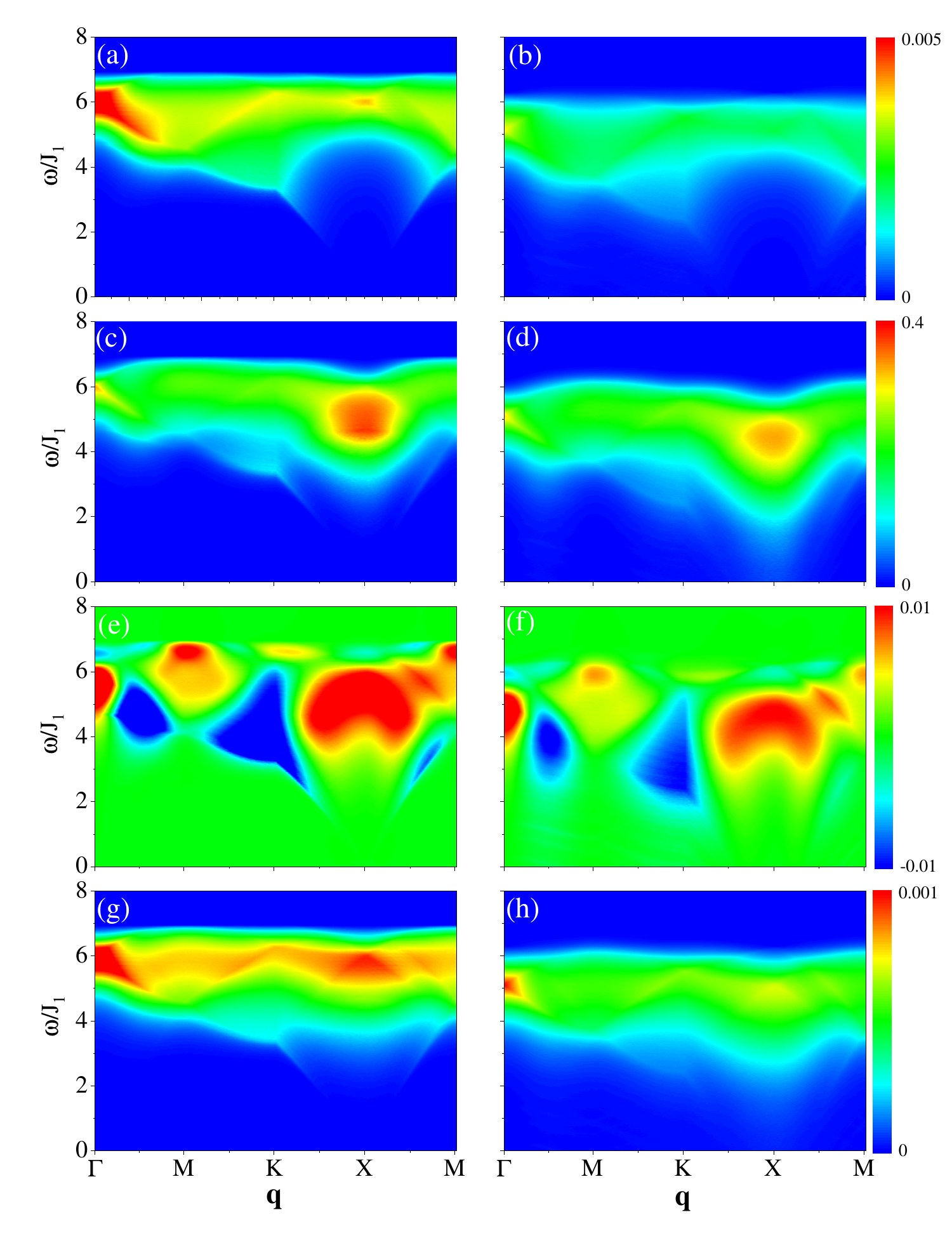}
\caption{Three-magnon RIXS intensity at $\mathcal{O}(\text{UCL}[1])$ and $\mathcal{O}(\text{UCL}[2])$. (a)–(b) $I^{(1,S^3,1)}_{\mathrm{3m}}(\qq,\omega)$ [Eq.~\eqref{eq:i33m}]. (c)–(d) $I^{(2_{jl},S^3,1)}_{\mathrm{3m}}(\qq,\omega)$ [Eq.~\eqref{eq:i43m}]. (e)–(f) $I^{([2_{ij},2_{jl}],S^3,1)}_{\mathrm{3m}}(\qq,\omega)$ [Eq.~\eqref{eq:i343m}]. (g)–(h) Total $S^3$ order three-magnon intensity $I^{S^3,\text{tot}}_{\mathrm{3m}}(\qq,\omega)$ [Eq.~\eqref{eq:itot3m}]. The left panels correspond to $(J_2,J_3,J_c) = 0$, while the right panels use $(J_2,J_3,J_c) = (0.05,0.05,0.4)J_1$. In panel (a), $I^{(1,S^3,1)}_{\mathrm{3m}}(\qq,\omega)$ exhibits a broad-band high-energy intensity with an upper band around $6.9J_1$, while panel (c) highlights intensity peaks around the $\mathrm{X}$ point, indicating enhanced three-magnon scattering in $I^{(2_{jl},S^3,1)}_{\mathrm{3m}}(\qq,\omega)$. The total intensity $I^{S^3,\text{tot}}_{\mathrm{3m}}(\qq,\omega)$ in panel (g) features broad-band peaks at high energy, and compared to panel (a), the higher-order UCL expansion enhances intensity peaks near the $\mathrm{K}$ and $\mathrm{X}$ points, reflecting six-spin correlations. Nonzero intensity peaks appear at $\Gamma$ point in all six-spin correlation spectra. With finite frustration parameters (panels (b), (d), (f), and (h)), the spectra shift to lower energies while retaining their overall spectral patterns.
}
\label{fig:fig11}
\end{figure}

In Fig.~\ref{fig:fig11}, we present the three-magnon RIXS intensity for different contributions: $I^{(1,S^3,1)}_{\mathrm{3m}}(\qq,\omega)$ [Eq.~\eqref{eq:i33m}] in (a) and (b), $I^{(2_{jl},S^3,1)}_{\mathrm{3m}}(\qq,\omega)$ [Eq.~\eqref{eq:i43m}] in (c) and (d), $I^{([2_{ij},2_{jl}],S^3,1)}_{\mathrm{3m}}(\qq,\omega)$ [Eq.~\eqref{eq:i343m}] in (e) and (f), and the total three-magnon RIXS intensity $I^{S^3,\text{tot}}_{\mathrm{3m}}(\qq,\omega)$ [Eq.~\eqref{eq:itot3m}] in (g) and (h). 
Overall, the figure shows that the three-magnon RIXS spectra in the six-spin correlation function exhibits a broad high-energy band structure, with intensity peaks that do not shift to the one-magnon excitation region. Panel (a) presents the three-magnon intensity $I^{(1,S^3,1)}_{\mathrm{3m}}(\qq,\omega)$ in $\mathcal{O}(\text{UCL}[1])$, with the upper band reaching approximately $6.9 J_1$. Panel (c) highlights the intensity of $I^{(2_{jl},S^3,1)}_{\mathrm{3m}}(\qq,\omega)$, which features a broad high-energy band with pronounced peaks around the $\mathrm{X}$ point. The cross-term contribution in panel (e) exhibits negative intensity, similar to the two-magnon cross-term contribution shown in Fig.~\ref{fig:fig6}(e). Panel (g) presents the total three-magnon intensity (overall all positive), primarily arising from the $\mathcal{O}(\text{UCL}[1])$ contribution, with a minor modification from the $\mathcal{O}(\text{UCL}[2])$ term. Compared to panel (a), the inclusion of higher-order UCL expansion enhances and localizes intensity peaks around the $\mathrm{K}$ and $\mathrm{X}$ points. With finite frustration parameters $(J_2,J_3,J_c)$ in panels (b), (d), (f), and (h), the broad-band spectrum shifts to a lower energy while retaining its overall spectral features.

The broad high-energy band feature of the three-magnon RIXS spectra in Fig.~\ref{fig:fig11} aligns with predictions from previous spin wave theory studies~\cite{ament2010strong,pal2023theoretical}.  However, compared to the $\mathcal{O}(1/S)$ and $\mathcal{O}[\text{UCL}(0)]$ three-magnon spectrum (displayed in Fig.~\ref{fig:fig9}), the $S^3$ order intensity $I^{S^3,\text{tot}}_{\mathrm{3m}}(\qq,\omega)$ will be scaled by $\xi^2/\Gamma^2$ and $\xi^4/\Gamma^2$. Thus it will be too weak to be experimentally detectable for $S=1/2$. To further consider the trimagnon spectra, one needs to incorporate perturbation vertices in the Hamiltonian and employ the ladder diagram approximation $T$-matrix formalism and the Faddeev equation groups~\cite{PhysRevA.107.042804}. However, as mentioned earlier, this approach is hindered by the presence of Faddeev spurious states, which contribute to at least two-thirds of the spectroscopic solutions~\cite{PhysRevC.63.034313}. Given the difficulties, rather than relying on perturbative vertex corrections within spin wave theory~\cite{verresen2018quantum}, we propose that the coherence of the three-magnon states can be effectively captured via the SBMFT approach.

\subsubsection{SBMFT result of six-spin correlation\label{subsubsec:4spi}}
\begin{figure*}[hbtp]
\includegraphics[width=\linewidth,scale=1.00]{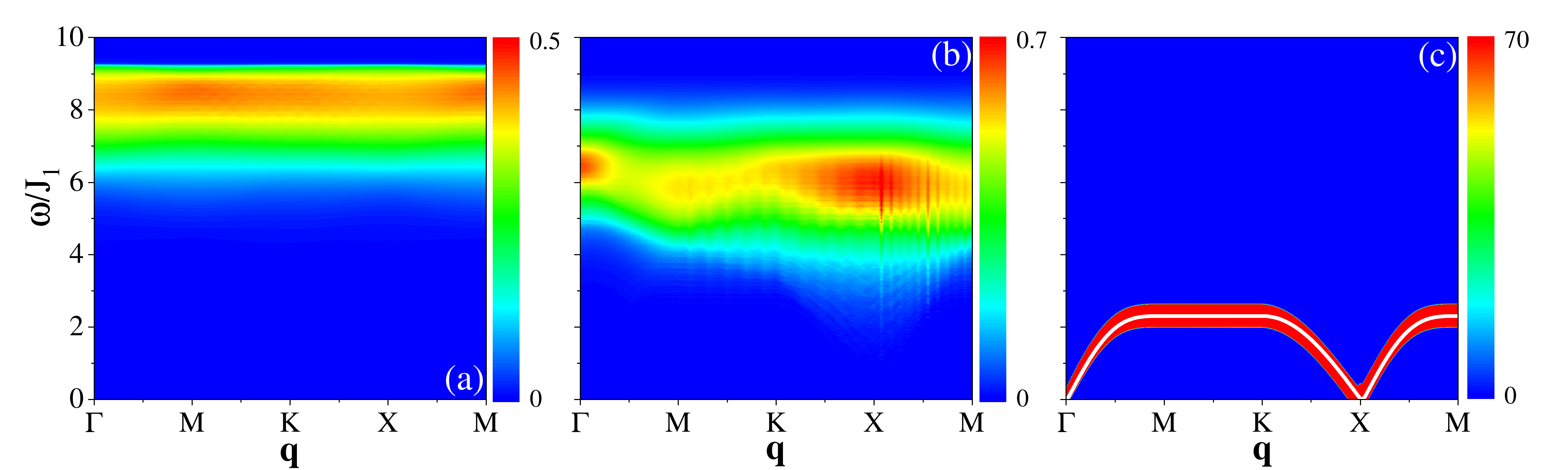}
\caption{(a) Mean-field four-spinon DOS $D_{\mathrm{4s}}(\qq,\omega)$ [Eq.~\eqref{eq:d4s}]. There is a continuum part contribution $I^c_{\mathrm{4s}}(\qq,\omega)$ [Eq.~\eqref{eq:i4sc}] (b) and a singular part contribution $I^s_{\mathrm{4s}}(\qq,\omega)$ [Eq.~\eqref{eq:i4ss}] (c). Panel (b) shows a broad spectral band ranging from $4J_1$ to $8J_1$, with sharp peaks around $6J_1$. The sharp peaks recover the main feature of the six-spin correlation spectrum, reflecting the three-magnon response, as shown in Fig.~\ref{fig:fig11}(g). In panel (c), a broad single-magnon curve (indicated by the red color band) is observed. This represents the spectroscopic behavior of mean-field four-spinon BEC, serving as a RIXS signature of the spinon condensate (Higgs phase) in the non-spin-conserving channel. For reference, compare the calculated spectrum with the sharp white line which displays the $1/S$ corrected single-magnon dispersion curve. As explained in the main text, except for $J_1$, all other parameters $(J_2,J_3,J_c)$ are set to zero.
}
\label{fig:fig12}
\end{figure*}
Within the context of SBMFT, spin-flip interactions alter the bosonic occupation states between two flavors of boson. This leads to boson-entangled excitations that either condense at $\pm \frac{\mathbf{Q}}{2}$ or confine for the rest of the momenta in the BZ. In the scattering operators for the three-spin-flip process in the Schwinger boson representation Eq.~\eqref{eq:o4s}, we apply a mean-field Ansatz [Eq.~\eqref{mfa}] to evaluate the spin-spin correlation term $(\hat{\bs}_i\cdot \hat{\bs}_j)_{MF}$ as a two-spinon contribution. The presence of an additional spin-flip operator $S_i^x$, introduces two additional entangled spinons. The resulting quasiparticle generates a mean-field four-spinon continuum response. The mean-field four-spinon $(\mathrm{4s})$ DOS $D_{\mathrm{4s}}(\qq,\omega)$ is given by 
\begin{align}
\notag&D_{\mathrm{4s}}(\qq,\omega) \\
   &= \frac{1}{N^3}\sum_{\substack{ \kk,\pp,\pp'\neq \pm\frac{\bq}{2} \\ \kk,\pp,\pp'\in BZ}} \delta\left(\omega-\omega^s_{\pp}-\omega^s_{\pp'}-\omega^s_{\kk+\qq}-\omega^s_{\kk+\pp+\pp'}\right).\label{eq:d4s}\end{align} 
The mean-field four-spinon continuum $\mathrm{(4sc)}$ RIXS intensity $I^c_{\mathrm{4s}}(\qq,\omega)$ and the mean-field four-spinon singular $\mathrm{(4ss)}$ RIXS intensity $I^s_{\mathrm{4s}}(\qq,\omega)$, computed using the scattering operators of Eq.~\eqref{eq:o4s}, are given by
   \begin{align}
\notag&I^c_{\mathrm{4s}}(\qq,\omega) \\
&\propto \frac{1}{N^3}\sum_{\substack{ \kk,\pp,\pp'\neq \pm \frac{\bq}{2}, \\ \kk,\pp,\pp'\in BZ}} |\mathrm{f_{4sc}}|^2 \delta\left(\omega-\omega^s_{\pp}-\omega^s_{\pp'}-\omega^s_{\kk+\qq}-\omega^s_{\kk+\pp+\pp'}\right)\label{eq:i4sc}, \\
\notag&I^s_{4s}(\qq,\omega) \propto (N m_0)^{3} \left\{ |\mathrm{f_{4ssp}}|^2 \delta\left(\omega-\omega^s_{\frac{\bq}{2}+\qq}\right) \right. \\
   &\quad + \left. |\mathrm{f_{4ssm}}|^2 \delta\left(\omega-\omega^s_{-\frac{\bq}{2}+\qq}\right) \right\}.\label{eq:i4ss}
\end{align}

Figure~\ref{fig:fig12} shows the DOS, the continuum part, and the singular part of the mean-field four-spinon RIXS intensity. In Fig.~\ref{fig:fig12}(a), the four-spinon DOS acquires a maximum around $4(2.3J_1)=9.2J_1$. In Fig.~\ref{fig:fig12}(b), a broad spectral band appears in the energy range from $4J_1$ to $8J_1$, with sharp peaks centered around $6J_1$. This spectrum reflects the six-spin correlation in the continuum part covering the results of the three-magnon RIXS response up to $\mathcal{O}(\text{UCL}[2])$ in Fig.~\ref{fig:fig11}(g). Fig.~\ref{fig:fig12}(c) shows that the BEC of the mean-field four-spinon would behave as a broad single-magnon dispersion curve in the NSC channel.

Next, we will discuss the observed satellite intensity peak in the vicinity of single-magnon~\cite{PhysRevLett.100.097001,PhysRevB.103.L140409,PhysRevLett.105.157006}. For the $L_3$-edge RIXS spectra, it has been proposed that this feature can be attributed to a three-magnon response arising from a one-to-three magnon hybridization process by analyzing polarization details~\cite{PhysRevB.85.064421, Igarashi_2015_effect}. Furthermore, in square lattice INS calculations, the one-to-three magnon hybridization process has been analyzed using continuous similarity transformation (CST) methods~\cite{PhysRevLett.115.207202,10.21468/SciPostPhys.4.1.001}. These studies suggest that the corresponding three-magnon continuum loses significant spectral weight and shifts to lower energies due to Higgs resonance. The satellite intensity of the single-magnon peak in INS has been interpreted to arise from the interacting three-magnon excitations leading to Higgs resonance~\cite{PhysRevLett.115.207202,10.21468/SciPostPhys.4.1.001}. A similar Higgs mechanism can also occur in our mean-field four-spinon singular part of the direct $L_3$-edge RIXS spectrum, see Fig.~\ref{fig:fig12}(c). In this case, spinon condensation drives a Higgs phase~\cite{PhysRevB.98.184403,PhysRevB.73.075119}, where the gauge bosons acquire a finite mass via the Higgs mechanism. The Higgs and confinement phases are compatible~\cite{PhysRevD.19.3682}. This leads to a mean-field four-spinon condensate spectrum in the RIXS NSC channel, which comes from the three-spin-flip scattering process within the SBMFT framework [Eq.~\eqref{eq:o4s}], as shown in Fig.~\ref{fig:fig12}(c). Therefore, within the scope of SBMFT interpretation, we find that an alternative viewpoint exists where the satellite intensity peak of the single-magnon spectrum can also arise from quantum fluctuations, quantum entanglement of the RVB phase, and Higgs mechanism of condensed spinons. Finally, we would like to state that it is possible to use Quantum Monte Carlo to calculate the proportions of N\'eel and RVB state in the square lattice~\cite{doi:10.1073/pnas.0703293104}. Additionally, one can perform polarization analysis \cite{PhysRevB.85.064421} considering the sum rule in RIXS~\cite{ament2011resonant,PhysRevB.58.3741} within SBMFT to accurately calculate the exact proportion of the three-magnon and the multi-spinon satellite RIXS intensity. We leave these as future topics of investigation.

\section{Conclusions}
Resonant inelastic x-ray scattering can probe quantum excitations ranging from low to high energies in correlated magnets. Inspired by the puzzling satellite intensity peaks observed above the single-magnon dispersion at the $L_3$-edge, we computed the RIXS spectra of the single-magnon, the multimagnon, and the multispinon excitations of the effective spin-1/2 $\tilde{J}_1$-$\tilde{J}_2$-$\tilde{J}_3$ Heisenberg model. We propose a RIXS mechanism inspired by RVB physics. We demonstrate that the RIXS features predicted by $1/S$-corrected spin-wave theory with higher order UCL expansion can be reproduced using SBMFT-interpreted RIXS operator matrix elements which incorporate RVB fluctuations in RIXS. Our analysis reveals that both three-magnon and condensed spinon excitations (in the Higgs phase) contribute to the satellite intensity peak in the neighborhood of the single magnon $L_3$-edge RIXS spectrum. Crucially, we show that these features arise from spectroscopic signatures of quantum fluctuations, quantum entanglement, and the Higgs mechanism. Our calculations highlight the interplay between magnetism and quantum many-body effects in frustrated antiferromagnets, while demonstrating how SBMFT can account for experimental data. 

\section*{Acknowledgment}

We would like to thank Ke-Jin Zhou, Umesh Kumar, Peng Ye, Junli Li, Muwei Wu, Meiyu He, Jia-Zheng Ma and Zijian Xiong for helpful discussions. Thanks to the Diamond Light Source for sharing unpublished experimental data. K.Y.Q. and D.X.Y. are supported by NKRDPC-2022YFA1402802, NSFC-92165204, NSFC-12494591, Guangdong Provincial Key Laboratory of Magnetoelectric Physics and Devices (No. 2022B1212010008), Research Center for Magnetoelectric Physics of Guangdong Province (2024B0303390001), and Guangdong Provincial Quantum Science Strategic Initiative (GDZX2401010).

\section*{Data availability}
The data that support the findings of this article are not publicly available. The data are available from the authors upon reasonable request.

\appendix

\section{Spin wave formalism and $1/S$-interacting Hamiltonian}
\label{app:LSWT+1/S}
In this appendix, we describe the HP, Fourier, and Bogoliubov transformations. We also outline the derivation of the $1/S$-interacting Hamiltonian.

\textit{Holstein-Primakoff transformation.}
The standard HP transformation of the spin flip operators $S^{\pm}$ and $S^z$ up to $\mathcal{O}(1/S)$ is given by
\begin{subequations}
\begin{eqnarray}
S^+_{Ai}&=&\sqrt{2S}f_i(n_i)a_i,\quad S^+_{Bj}=\sqrt{2S}b^\dagger_j f_j(n_j),\\
S^-_{Ai}&=&\sqrt{2S}a^\dagger_i f_i(n_i),\quad S^-_{Bj}=\sqrt{2S}f_j(n_j)b_j,\\
S^z_{Ai}&=&S-n_i,\quad S^z_{Bj}=-S+n_j,
\end{eqnarray}
\end{subequations} where $f_l(n_l)= \left(1-\frac{n_l}{4S}\right)$, $l=i,j,n_i=a_i^\dagger a_i$, and $n_j=b_j^\dagger b_j$. The $A$ and $B$ sublattice single magnon creation (annihilation) operators are denoted by $a^{\dagger}_i (a_i)$ and $b^{\dagger}_j (b_j)$, respectively. 

\textit{Fourier transformation.}
The sublattice Fourier transforms are defined as
\begin{equation}
a_i=\sqrt{\frac{2}{N}}\sum_{\kk}e^{i\kk \cdot \br_i} a_\kk,~~~
b_j=\sqrt{\frac{2}{N}}\sum_{\kk}e^{i\kk \cdot \br_j} b_\kk, 
\end{equation} where $\br_i$ and $\br_j$ indicate the location of the magnetic sites, $N$ is the number of lattice sites, and the momenta $\kk$ is defined in the MBZ. 

\textit{Bogoliubov transformation and 1/S-interacting Hamiltonian.}
The transformation matrix between the original $(a_\kk, b_\kk)$ and the Bogoliubov bosons $(\alpha_\kk,\beta_\kk)$ is given by ~\cite{van1980note} 
\begin{align}
       	\begin{pmatrix}
	     \alpha_\kk \\
        \beta_{-\kk}^\dagger
	\end{pmatrix}=	
 \begin{pmatrix}
	     u_\kk & v^\ast_{-\kk} \\
        v_\kk & u^\ast_{-\kk}
	\end{pmatrix}
 	\begin{pmatrix}
	     a_\kk \\
        b_{-\kk}^\dagger
	\end{pmatrix},
   \end{align} where the Bogoliubov coefficients $u_\kk$ and $v_\kk$ are 
\begin{align}
     u_\kk &= \left(\frac{1+\epsilon_\kk}{2\epsilon_\kk}\right)^\frac{1}{2},
     v_\kk = \text{sign}(\gamma_\kk)\left(\frac{1-\epsilon_\kk}{2\epsilon_\kk}\right)^\frac{1}{2}=x_\kk u_\kk. \notag \\
     & 
\end{align}
The \text{sign}$(x)$ function takes the value $+1$ when $x>0$ and $-1$ when $x<0$. The final Bogoliubov transformed and diagonalized Hamiltonian $\mathcal{H}_0 =H_0/4J_1S$ is \begin{equation}
\mathcal{H}_0=\sum\limits_\kk\kappa_\kk(\epsilon_\kk-1)+\sum\limits_\kk\kappa_\kk\epsilon_\kk(\alpha_k^\dagger\alpha_k+\beta_{-k}^\dagger\beta_{-k}).
\end{equation} The first term is the zero-point energy and the second term represents the excitation energy of the magnons. The functions $\kappa_\kk$ and $\epsilon_\kk$ are given by \begin{subequations}
\begin{flalign}
\kappa_{\kk}&=R_1-R_2\left(1-\gamma_{2}(\kk)\right)-R_3\left(1-\gamma_{3}(\kk)\right), \label{eq:first}\\
\epsilon_{\bm{\kk}}&=\sqrt{1-\gamma_{\kk}^2},~~~\gamma_{\kk}= R_1 \frac{\gamma_1(\kk)}{\kappa_{\kk}},\\
\gamma_{1}(\kk)&= \frac{1}{2}\left[\cos(k_x)+\cos(k_y)\right], \\
\gamma_{2}(\kk)&=\cos(k_x)\cos(k_y),\\
\gamma_{3}(\kk)&=\frac{1}{2}\left[\cos(2k_x)+\cos(2k_y)\right], \label{eq:last}
\end{flalign}
\end{subequations} where $R_l=\tilde{J_l}/J_1,l=1,2,3$. 

Next, we state the expressions for the vertex functions of the effective $\tilde{J}_1-\tilde{J}_2-\tilde{J}_3$ model up $\mathcal{O}(1/S)$. The vertices capture the magnon-magnon interactions which are a manifestation of the many-body spin-wave scattering processes. These vertices represent the lowest-order self-energy correction of the two-magnon propagators~\cite{ canali1992theory}. The first term in Eq.~\eqref{eq:Svertex1} is known as the Oguchi correction which represents the one-body approximation of the quartic interaction terms up to $\mathcal{O}(1/S)$. Thus, the complete $\mathcal{O}(1/S)$ perturbative Hamiltonian $\mathcal{H}_1 =H_1/4J_1S$ is given by \begin{align}
\label{eq:Svertex1}
\notag \mathcal{H}_1&=\frac{1}{2S}\sum\limits_\kk [ A_\kk(\alpha^\dagger_k 	
\notag \alpha_k+\beta^\dagger_{-k} \beta_{-k})+B_\kk(\alpha^\dagger_k 	
\notag \beta_{-k}+ \text{H.c.} )]\\
\notag&+\frac{1}{2SN}\sum_{1234}\delta_\bg(1+2-3-4)u_1u_2u_3u_4\\
\notag&[\alpha_1^\dagger\alpha^\dagger_2\alpha_3\alpha_4 
	V^{(1)}_{1234}+\beta^\dagger_{-3}\beta^\dagger_{-4}\beta_{-1}\beta_{-2}V^{(2)}_{1234}+\\
	&4\alpha^\dagger_{1}\beta^\dagger_{-4}\beta_{-2}\alpha_3V^{(3)}_{1234}+(2\alpha^\dagger_{1}\beta_{-2}\alpha_3\alpha_4V^{(4)}_{1234}+\\
 \label{eq:Svertex2}
\notag&2\beta^\dagger_{-4}\beta_{-1}\beta_{-2}\alpha_3V^{(5)}_{1234}+\alpha^\dagger_1\alpha^\dagger_2\beta^\dagger_{-3}\beta^\dagger_{-4}V^{(6)}_{1234}+\text{H.c.})],
\end{align} 
The momenta $\kk_1,\kk_2, \kk_3...$ are abbreviated as 1,2,3... and $\bg$ is the reciprocal lattice vector. To derive the above expression we used Eqs.~\eqref{eq:first}-\eqref{eq:last} and the Oguchi correction coefficients 
\begin{subequations}
\begin{eqnarray}
&A_\kk=A_1\frac{1}{\kappa_\kk \epsilon_\kk}[\kappa_\kk-\gamma^2_1(\kk)]+A_2\frac{1}{ \epsilon_\kk}[1-\gamma_2(\kk)],\nonumber\\
		&+A_3\frac{1}{ \epsilon_\kk}[1-\gamma_3(\kk)]\\
	&B_\kk=\frac{\gamma_1(\kk)}{\kappa_\kk \epsilon_\kk}\Big\{[1-\gamma_2(\kk)]B_1+[1-\gamma_3(\kk)]B_2\Big\}. 
\end{eqnarray}
\end{subequations}
The symbols $A_1, A_2, ...$ are the Hartree-Fock average terms defined as \begin{flalign}
&A_1=R_1\frac{2}{N}\sum_\pp\frac{1}{\epsilon_\pp}[\frac{\gamma^2_1(\pp)}{\kappa_\pp}+\epsilon_\pp-1],\\ 
&A_n=R_n\frac{2}{N}\sum_\pp\frac{1}{\epsilon_\pp}[1-\epsilon_\pp-\gamma_n(\pp)],\\
&B_m=R_{m+1}\frac{2}{N}\sum_\pp\frac{1}{\epsilon_\pp}[\gamma_{m+1}(\pp)-\frac{\gamma^2_1(\pp)}{\kappa_\pp}],
\end{flalign} where $n=2,3$ and $m=1,2$. The explicit expressions for the vertex functions $V^{(1)}_{1234},V^{(2)}_{1234},V^{(3)}_{1234}...$ are 
\begin{widetext}
\begin{subequations}
\label{app:vertex}
\begin{eqnarray}
V^{(1)}_{1234}&=&-[\gamma_1(4-2)x_2x_4+\gamma_1(4-1)x_1x_4+\gamma_1(3-2)x_2x_3 +\gamma_1(3-1)x_1x_3]+\frac{1}{2}[\gamma_1(1)x_1+\gamma_1(2)x_2+\gamma_1(3)x_3+\gamma_1(4)x_4 \nonumber\\
  &+&\gamma_1(4-1-2)x_1x_2x_4+\gamma_1(3-1-2)x_1x_2x_3+\gamma_1(3+4-2)x_2x_3x_4+\gamma_1(3+4-1)x_1x_3x_4]\nonumber\\
  &+&(R_2 F^2_{1234}+R_3 F^3_{1234})[1+\mathrm{sign}(\gamma_\bg)x_1x_2x_3x_4],\\
V^{(2)}_{1234}&=&-[\gamma_1(4-2)x_1x_3+\gamma_1(4-1)x_2x_3+\gamma_1(3-2)x_1x_4+\gamma_1(3-1)x_2x_4]+\frac{1}{2}[\gamma_1(1)x_2x_3x_4+\gamma_1(2)x_1x_3x_4\nonumber\\
&+&\gamma_1(3)x_1x_2x_4+\gamma_1(4)x_1x_2x_3+\gamma_1(4-1-2)x_3+\gamma_1(3-1-2)x_4+\gamma_1(3+4-2)x_1+\gamma_1(3+4-1)x_2]\nonumber\\
&+&(R_2 F^2_{1234}+R_3 F^3_{1234})[x_1x_2x_3x_4+\mathrm{sign}(\gamma_\bg)],\\
V^{(3)}_{1234}&=&-[\gamma_1(4-2)+\gamma_1(4-1)x_1x_2+\gamma_1(3-2)x_3x_4+\gamma_1(3-1)x_1x_2x_3x_4]+\frac{1}{2}[\gamma_1(1)x_1x_2x_4+\gamma_1(2)x_4\nonumber\\
&+&\gamma_1(3)x_2x_3x_4+\gamma_1(4)x_2
+\gamma_1(4-1-2)x_1+\gamma_1(3-1-2)x_1x_3x_4
+\gamma_1(3+4-2)x_3+\gamma_1(3+4-1)x_1x_2x_3]\nonumber\\
&+&(R_2 F^2_{1234}+R_3 F^3_{1234})[x_2x_4+\mathrm{sign}(\gamma_\bg)x_1x_3],\\
V^{(4)}_{1234}&=&\gamma_1(4-2)x_4+\gamma_1(4-1)x_1x_2x_4+\gamma_1(3-2)x_3+\gamma_1(3-1)x_1x_2x_3-\frac{1}{2}[\gamma_1(1)x_1x_2+\gamma_1(2)+\gamma_1(3)x_2x_3\nonumber\\
&+&\gamma_1(4)x_2x_4+\gamma_1(4-1-2)x_1x_4+\gamma_1(3-1-2)x_1x_3+\gamma_1(3+4-2)x_3x_4+\gamma_1(3+4-1)x_1x_2x_3x_4]\nonumber\\
&-&(R_2 F^2_{1234}+R_3 F^3_{1234})[x_2+\mathrm{sign}(\gamma_\bg)x_1x_3x_4],\\
V^{(5)}_{1234}&=&\gamma_1(4-2)x_1+\gamma_1(4-1)x_2+\gamma_1(3-2)x_1x_3x_4+\gamma_1(3-1)x_2x_3x_4-\frac{1}{2}[\gamma_1(1)x_2x_4+\gamma_1(2)x_1x_4\nonumber\\
&+&\gamma_1(3)x_1x_2x_3x_4+\gamma_1(4)x_1x_2+\gamma_1(4-1-2)+\gamma_1(3-1-2)x_3x_4+\gamma_1(3+4-2)x_1x_3+\gamma_1(3+4-1)x_2x_3]\nonumber\\
&-&(R_2 F^2_{1234}+R_3 F^3_{1234})[x_1x_2x_4+\mathrm{sign}(\gamma_\bg)x_3],\\
V^{(6)}_{1234}&=&-[\gamma_1(4-2)x_2x_3+\gamma_1(4-1)x_1x_3+\gamma_1(3-2)x_2x_4+\gamma_1(3-1)x_1x_4]+\frac{1}{2}[\gamma_1(1)x_1x_3x_4+\gamma_1(2)x_2x_3x_4\nonumber\\
&+&\gamma_1(3)x_4+\gamma_1(4)x_3+\gamma_1(4-1-2)x_1x_2x_3+\gamma_1(3-1-2)x_1x_2x_4+\gamma_1(3+4-2)x_2+\gamma_1(3+4-1)x_1]\nonumber\\
&+&(R_2 F^2_{1234}+R_3 F^3_{1234})[x_3x_4+\mathrm{sign}(\gamma_\bg)x_1x_2].
\end{eqnarray}

\end{subequations}
Here, the \text{sign}$(\gamma_\bg)$ is a consequence of the umklapp process~\cite{PhysRevB.45.10131}. The presence of $J_2$ and $J_3$ generates vertex terms $F^{2,3}_{1234}$ given by \begin{equation}
F^l_{1234}=\frac{1}{2}[\gamma_l(2-4)+\gamma_l(1-4)+\gamma_l(2-3)+\gamma_l(1-3)-\gamma_l(1)-\gamma_l(2)-\gamma_l(3)-\gamma_l(4)],~
l=2,3.
\end{equation}

\end{widetext} 
\section{Schwinger boson mean field theory} \label{app:sbmft} 

\subsection{Mean-field Hamiltonian}
To perform SBMFT on the antiferromagnetic square lattice Heisenberg model, we define the singlet bond operator $\hat{A}_{ij}=\frac{1}{2}[a^s_ia^s_j+b^s_ib^s_j]$ to preserve translational invariance in the antiferromagnetic phase~\cite{auerbach2012interacting,pires2021theoretical}. 
One could consider triplet excitations in the square lattice for further study~\cite{ghioldi2016rvb}. For simplicity, we just treat the singlet RIXS case since this already allows for an explanation for the satellite intensity. The SBMFT Hamiltonian can be expressed in terms of the bond operator as
\begin{equation}
    \hat{H}=J_1\sum_{\langle i,j\rangle}\left(S^2-2\hat{A}^\dagger_{ij}\hat{A}_{ij}\right).
\end{equation} 
Next, we invoke the Lagrange multiplier $\lambda$ which implements the constraint $a_i^{s\dagger} a^s_i+ b_i^{s\dagger} b^s_i=2S$. Using the mean-field decoupling ansatz
\begin{align}
\langle\hat{A}_{ij}^\dagger \hat{A}_{ij}\rangle=\hat{A}_{ij}^\dagger \langle\hat{A}_{ij}\rangle+ \langle\hat{A}_{ij}^\dagger\rangle\hat{A}_{ij}-\langle\hat{A}_{ij}^\dagger\rangle\langle\hat{A}_{ij}\rangle,\label{mfa}
\end{align}
where $\langle\hat{A}_{ij}\rangle=\langle\hat{A}_{ij}^\dagger\rangle=A_\delta$, we obtain the mean-field Hamiltonian $\hat{H}_{MF}$ as
		\begin{align}
 \notag \hat{H}_{MF}&=-2J_1\sum_{i,j}\left[A_{\delta} \left(\hat{A}_{ij}^\dagger +H.c.\right)\right]+NJA_{\delta}^2\nonumber\\
 &+\lambda\sum_i\Big(a^{s\dagger} _ia^s_i+b^{s\dagger} _ib^s_i-2S\Big).
 \label{eq:hmf}
	\end{align}
Note, Eq.~\eqref{eq:hmf} exhibits an emergent gauge redundancy~\cite{PhysRevB.109.134409}.  As we discuss later, the conceptual possibility of fluctuations arising from this intrinsic gauge field is crucial to explaining our proposed RIXS bond spin-flip mechanism at the $L$-edge. 

To examine the stability of the mean-field solutions one can consider the effects of gauge fluctuation. The customary way to do this is through a large $N$ expansion and use a perturbative treatment to evaluate fluctuations around mean-field solutions~\cite{PhysRevB.38.316, auerbach2012interacting, PhysRevB.105.224404}. However, in this approach, one cannot obtain a confined spinon phase and handle all possible fluctuations adequately~\cite{lacroix2011introduction}. Thus, 
as mentioned earlier instead of a perturbative treatment, the invariant gauge group (IGG) approach is more appropriate~\cite{PhysRevB.65.165113}. In this technique, the fluctuation modes around the mean-field solutions can be captured via gauge fluctuations, whose nature is determined by the IGG~\cite{jacobsen2010exact}. In our case the spinons are confined, thereby protecting the parent antiferromagnetic phase while preserving gauge fluctuations. The presence of such a fluctuation scenario validates our RIXS mechanism scheme at the $L$-edge. 

In SBMFT, one site spin-flip  involves changing the occupation number of two flavors of bosons in its primitive cell. Thus, we use a $4\times 4$ Bogoliubov transformation matrix defined as 
    \begin{align}
     	\begin{pmatrix}
	     a^s_\kk \\
      b^s_\kk \\
      a^{s\dagger}_{-\kk}\\
        b^{s\dagger}_{-\kk}
	\end{pmatrix}=
 \begin{pmatrix}
	     u^s_\kk &0& v^{s\ast}_{-\kk} &0\\
      0&  u^s_\kk &0 &v^{s\ast}_{-\kk}\\
        v^s_\kk &0& u^{s\ast}_{-\kk}&0 \\
       0 &v^s_\kk &0& u^{s\ast}_{-\kk}
	\end{pmatrix}
        	\begin{pmatrix}
	     \alpha^s_\kk \\
      \beta^s_{\kk}\\
      \alpha^{s\dagger}_{-\kk}\\
        \beta^{s\dagger}_{-\kk}
	\end{pmatrix},	
   \end{align}
where the Bogoliubov coefficients are given by
 \begin{align}
     u^s_\kk=\sqrt{\frac{\lambda+\omega^s_\kk}{2\omega^s_\kk}}, \quad v^s_\kk=i\cdot \text{sign}(\gamma_\kk^A)\sqrt{\frac{\lambda-\omega^s_\kk}{2\omega^s_\kk}}.
 \end{align}

After performing the Bogoliubov transformation, the diagonalized mean-field Hamiltonian is 
	\begin{align}		
 \hat{H}^s_{MF}=\sum_k\omega^s_\kk\Big(\alpha_\kk^{s\dagger} \alpha^s_\kk+\beta_{-\kk}^{s\dagger} \beta^s_{-\kk}+1\Big)+E_g,
	\end{align}
where $\omega^s_\kk=\sqrt{\lambda-z\lvert A_{\delta}\gamma^A_\kk\rvert^2}$ is the mean-field spinon dispersion relation with $\gamma^A_\kk=i\cdot\frac{2}{z}\sum\limits_{\delta>0} \sin (\kk \cdot \delta)$ and $E_g=\frac{1}{2}NzA_\delta^2-2N\lambda (S+\frac{1}{2})$ is the zero-mode energy~\cite{lacroix2011introduction}. Under the thermodynamic limit $N\rightarrow \infty$ and $T\rightarrow 0$, the gap at $\pm\frac{\bq}{2}=\pm( \frac{\pi}{2},\frac{\pi}{2})$ will disappear and BEC is reached, leading to solutions for magnetization and $A_{\delta}$~\cite{auerbach2012interacting}. The self-consistent Eqs.~\eqref{eq:spinmf} are obtained by minimizing the free energy $F$, which is 
    \begin{align}\label{eq:free_F}
        F=k_B T \sum_{\mu  k} \ln\Big(1-e^{-\omega_{\mu  k}/k_B T}\Big)+E_g,\mu=\alpha^s,\beta^s
    \end{align}

\section{RIXS operator and scattering matrix elements}\label{app:rso}

In this appendix, we derive the RIXS operator matrix element expressions based on the conceptual picture of the local spin-flip scheme with $1/S$ corrected spin wave theory and the bond spin-flip RIXS scheme based on SBMFT. The UCL expansion provides a commonly used scheme for calculating the local spin-flip RIXS spectrum~\cite{PhysRevB.77.134428,PhysRevB.75.115118}. It expresses the spin excitation in the RIXS cross-section into spin-wave perturbative solutions and leads to various combinations of spin-flip operators~\cite{luo2014spectrum}. The $1/S$ expansion generates the single-magnon, two-magnon, and three-magnon excitations. Two (three) spin-flip operators describe the two (three) -magnon excitation. Single-magnon and three-magnon excitations are in the NSC channel ($\Delta S=1$). The two-magnon excitation exists in the SC channel ($\Delta S=0$). The general form of the $1/S$ corrected RIXS operator matrix element and operator has been outlined in the main text. In this appendix, building on [Eq.~\eqref{eq:oql}] and the perturbation order $(\mathcal{L},\mathcal{M},\mathcal{N})$ conventions outlined in Sec.~\ref{sec:rixssop}, we derive and state the explicit expressions for the RIXS operator matrix elements mentioned in the main text. Note, as mentioned in Sec.~\ref{subsec:EffectiveHeisenbergmodel}, the Bogoliubov transformed RIXS operators are derived under the assumption of a N\'{e}el ordered ground state.

\subsection{$1/S$ expansion RIXS operator matrix elements}\label{app:LSWT+1/Sso}
The zeroth-order single-magnon (denoted by 1m) RIXS operator matrix element is given by
\begin{equation}
\mathrm{O}^{(0,\sqrt{S},1)}_{\qq,\mathrm{1m}}=\textstyle{\langle \mathrm{1m}|}\sum\limits_ie^{i\qq \cdot \br_i}\hat{S}^x_ih_ih^\dagger_i|g\rangle, \label{eq:o0s1m}
\end{equation}
where $h_i$ stands for the core-hole annihilation operator on site $i$, $|g\rangle$ is the ground state, and $|\mathrm{1m}\rangle$ is the one-magnon excited state. Thus, $\mathrm{O}^{(0,\sqrt{S},1)}_{\qq,\mathrm{1m}}$ can be rewritten in terms of the Bogoliubov quasiparticles $\alpha^\dagger_\qq$ and $\beta^\dagger_\qq$ as  \begin{align}
\mathrm{O}^{(0,\sqrt{S},1)}_{\qq,\mathrm{1m}}=\textstyle{\langle \mathrm{1m}|}\sqrt{\frac{SN}{2}}\sum\limits_i(u_\qq-v_\qq)(\alpha^\dagger_{-\qq}+\beta_{-\qq}^\dagger)|g\rangle+H.c..
\end{align}
This is used to calculate the spectral weight of the single-magnon (see Fig.~\ref{fig:fig4}).

The first-order two-magnon (denoted by 2m) operator matrix element is given by the expression
\begin{align}
\mathrm{O}^{(1,S,0)}_{\qq,\mathrm{2m}}=\langle \mathrm{2m}|\sum_{i,j}e^{i\qq \cdot \br_i}h_i(\hat{\bs}_i \cdot \hat{\bs}_j)h_i^\dagger|g\rangle,\label{eq:o1s2m}
\end{align}
where $|\mathrm{2m}\rangle$ is the two-magnon excited state. We decompose this operator into the $A$ and $B$ sublattices before transforming it to the Bogoliubov version to obtain  
\begin{align}
\label{eq:o1s}
    &\mathrm{O}^{(1,S,0)}_{\qq,\mathrm{2m}}=\langle \mathrm{2m}|S\sum\limits_\kk \mathrm{f^{(1)}_{2m}}(\kk,\qq)\alpha_{\kk+\frac{\qq}{2}}^\dagger\beta^\dagger_{-\kk+\frac{\qq}{2}}|g\rangle+H.c..
\end{align} 
The two-magnon first order scattering matrix element $\mathrm{f^{(1)}_{2m}}(\kk,\qq)$ is 
\begin{eqnarray}
    &&\mathrm{f^{(1)}_{2m}}(\kk,\qq) =  -\left\{R_2\left[\gamma_2(\kk-\frac{\qq}{2})+\gamma_2(\kk+\frac{\qq}{2})-1-\gamma_2(\qq)\right]\right.\nonumber \\
    &&\left.+R_3\left[\gamma_3(\kk-\frac{\qq}{2})+\gamma_3(\kk+\frac{\qq}{2})-1-\gamma_3(\qq)\right]\right.\nonumber \\
    &&\left.+R_1[1+\gamma_1(\qq)]\right\}(u_{\kk+\frac{\qq}{2}}v_{\kk-\frac{\qq}{2}}+v_{\kk+\frac{\qq}{2}}u_{\kk-\frac{\qq}{2}})  \nonumber \\
     &&\left. +R_1\left[\gamma_1(\kk-\frac{\qq}{2})+\gamma_1(\kk+\frac{\qq}{2})\right](u_{\kk+\frac{\qq}{2}}u_{\kk-\frac{\qq}{2}}+v_{\kk+\frac{\qq}{2}}v_{\kk-\frac{\qq}{2}}). \right. \nonumber \\
\end{eqnarray}\label{app:msme} 
which is included in Eq.~\eqref{eq:i12m} for computing Figs.~\ref{fig:fig5}(c) and \ref{fig:fig5}(d).

The second-order correction of the UCL expansion creates a 4-spin-flip process. This operator matrix element is expressed as 
\begin{align}
    \mathrm{O}^{(2,S,0)}_{\qq,\mathrm
{2m}}=\langle \mathrm
{2m}|\sum\limits_{i,j,l}e^{i\qq \cdot \br_i}h_i(\hat{\bs}_i\cdot \hat{\bs}_j)(\hat{\bs}_i\cdot \hat{\bs}_l)h_i^\dagger|g\rangle.\label{eq:2ucl4s}
\end{align}
The 4-spin-flip process (analyzed within the HP boson formalism) can be simplified using the approximation~\cite{PhysRevB.77.134428,PhysRevB.106.L060406,pal2023theoretical}
\begin{equation}
(\hat{\bs}_i\cdot \hat{\bs}_j)(\hat{\bs}_i\cdot \hat{\bs}_l)\approx \frac{1}{4}\hat{\bs}_j\cdot \hat{\bs}_l- \frac{1}{2}\hat{\bs}_i\cdot \hat{\bs}_j\label{eq:4s},
\end{equation}
where $j\neq l$ with $j,l\in NN$ of $i$. This generates two second order terms whose UCL order will be labeled as $2_{jl}$ and $2_{ij}$. The RIXS operator matrix elements belonging to these terms are 
\begin{eqnarray}
\mathrm{O}^{(2_{jl},S,0)}_{\qq,\mathrm{2m}}=\langle \mathrm{2m}|\sum\limits_{i,j,l}e^{i\qq \cdot \br_i} h_i(\hat{\bs}_j\cdot \hat{\bs}_l) h_i^\dagger|g\rangle.	
\label{eq: o2sq2m}\\
\mathrm{O}^{(2_{ij},S,0)}_{\qq,\mathrm{2m}}=\langle \mathrm{2m}|\sum_{i,j}e^{i\qq \cdot \br_i}h_i(\hat{\bs}_i \cdot \hat{\bs}_j)h_i^\dagger|g\rangle,\label{eq:o2s2m} 
\end{eqnarray}
Following the $A$ and $B$ sublattice decomposition outlined previously, we obtain the Bogoliubov transformed version as   
\begin{equation}
\mathrm{O}^{(2_{jl},S,0)}_{\qq,\mathrm{2m}}=\langle \mathrm{2m}|S\sum\limits_\kk \mathrm{f^{(2_{jl})}_{2m}}(\kk,\qq)\alpha^\dagger_\kk\beta^\dagger_{-\kk+\qq}|g\rangle +H.c. . \label{eq:o2p2m}
\end{equation}
The second-order scattering matrix element $\mathrm{f^{(2_{jl})}_{2m}}(\kk,\qq)$ is given by the expression 
  \begin{eqnarray}
      &&\mathrm{f^{(2_{jl})}_{2m}}(\kk,\qq)=\left\{-6[\cos(q_x)+\cos(q_y)]+2[\cos(2k_x-q_x)\right.  \nonumber\\
&&\left.+\cos(2k_y-q_y)] +4[\cos(k_x)\cos(k_y-q_y)\right. \nonumber\\
&&\left.+\cos(k_y)\cos(k_x-q_x)]\right\} (u_\kk v_{\kk-\qq}+v_\kk u_{\kk-\qq}).
  \end{eqnarray}
Furthermore $\mathrm{f^{(2_{ij})}_{2m}}(\kk,\qq)=\mathrm{f^{(1)}_{2m}}(\kk,\qq)$. These equations appear in Eq.~\eqref{eq:i22m} in the main text. 

The three-magnon (denoted by 3m) continuum is generated from the $1/S$ expansion of the single-site spin-flip operator~\cite{PhysRevB.48.3264,PhysRevB.72.014403,PhysRevB.85.064421} and the three-spin-flip operators~\cite{ament2010strong,pal2023theoretical}. Note, as evident from the equations below, these contributions are at different levels of spin wave expansion. The zero-th order three-magnon RIXS operator matrix element is 
\begin{equation}
\mathrm{O}^{(0,S^{-\frac{1}{2}},1)}_{\qq,\mathrm{3m} }=\textstyle{\langle \mathrm{3m}|}\sum\limits_i e^{i\qq \cdot \br_i}\hat{S}^x_ih_ih^\dagger_i|g\rangle,\label{eq:o03m}
\end{equation}
with $|{\mathrm{3m}}\rangle$ denoting the excited states of three-magnon continuum. The spectroscopic contribution of the one-to-three magnon hybridization process in the RIXS intensity at $\mathcal{O}(1/S)$ is computed by first defining the following operators~\cite{PhysRevB.72.014403,PhysRevB.85.064421}
\begin{align}
Y_\alpha^+=[Y_\alpha^-] ^\dagger=[u_\kk S_A^+(\kk)+v_\kk S_B^+(\kk)]/(2S)^{\frac{1}{2}},\\
Y_\beta^+=[Y_\beta^-] ^\dagger=[v_\kk S_A^+(\kk)+u_\kk S_B^+(\kk)]/(2S)^{\frac{1}{2}},
\end{align} 
where $S_A^+(\kk)$ and $S_B^+(\kk)$ are the Fourier transformed versions of $S_{Ai}^+$ and $S_{Bj}^+$, respectively. The total contribution from the three-magnon terms at the zero-th order of UCL expansion can be expressed as 
 \begin{eqnarray}
&&Y_\alpha^++Y_\alpha^- +Y_\beta^++Y_\beta^-=\notag\\
&&\mathrm{f^{(\mathsf{a})}_{3m}}(\kk,\pp,\qq)\alpha^\dagger_{-\kk}\beta^\dagger_{\pp}\beta^\dagger_{\kk-\pp-\qq} \notag\\
&&+\text{sign}(\gamma_{\bold{G}})\mathrm{f^{(\mathsf{b})}_{3m}}(\kk,\pp,\qq)\beta^\dagger_{-\kk}\alpha^\dagger_{\pp}\alpha^\dagger_{\kk-\pp-\qq},
\end{eqnarray}
with the following scattering matrix element definitions
\begin{flalign}
 & \mathrm{f^{(\mathsf{a})}_{3m}}(\kk,\pp,\qq)=-u_\qq v_\kk u_\pp u_{\pp+\qq-\kk} +\text{sign}(\gamma_{\bold{G}})v_\qq u_\kk v_\pp v_{\pp+\qq-\kk}, \\
 &\mathrm{f^{(\mathsf{b})}_{3m}}(\kk,\pp,\qq)=u_\qq u_\kk v_\pp v_{\pp+\qq-\kk} -\text{sign}(\gamma_{\bold{G}})v_\qq v_\kk u_\pp u_{\pp+\qq-\kk}.
 \end{flalign}
where \text{sign}$(\gamma_\bg)$ is a consequence of the umklapp process~\cite{PhysRevB.45.10131}. The above expressions are used in Eq.~\eqref{eq:i1s3m} and calculate Figs.~\ref{fig:fig9}(c) and \ref{fig:fig9}(d). The first-order three-spin-flip RIXS operator matrix element and the Bogoliubov transformed equations are given by
\begin{eqnarray}
\mathrm{O}^{(1,S^{\frac{3}{2}},1)}_{\qq,\mathrm{3m}}=	\langle \mathrm{3m}|\sum\limits_{i,j}e^{i\qq \cdot \br_i}\hat{S}^x_ih_i(\hat{\bs}_i \cdot \hat{\bs}_j)h_i^\dagger|g\rangle,\label{eq:o13m}
\end{eqnarray}
\begin{eqnarray}
			&&\mathrm{O}^{(1,S^{\frac{3}{2}},1)}_{\qq,\mathrm{3m}}=\langle \mathrm{3m}|\frac{S^\frac{3}{2}}{\sqrt{2N}}\sum\limits_{\kk,\pp}\left[\mathrm{f^{(\mathsf{c})}_{3m}}(\kk,\pp,\qq)\alpha^\dagger_{\kk+\pp+\qq}\alpha^\dagger_{-\pp}\beta^\dagger_{-\kk} \right.\nonumber\\
+ &&\left.\mathrm{f^{(\mathsf{d})}_{3m}}(\kk,\pp,\qq)\alpha^\dagger_{\kk+\pp+\qq}\beta^\dagger_{-\pp}\beta^\dagger_{-\kk}\right]|g\rangle+H.c.,\label{eq:o13mb}
\end{eqnarray}
with the scattering matrix elements
\begin{eqnarray}
     &&\mathrm{f^{(\mathsf{c})}_{3m}}(\kk,\pp,\qq) = u_{\kk+\pp+\qq}v_\kk\left\{\left[R_1\gamma_1(\pp+\qq)+J^{23}_{\kk,\pp,\qq}\right]v_\pp
    \right. \nonumber \\
    &&\left.-(R_1+J^{23}_{\kk,\pp,\qq})u_\pp\right\}  +v_{\kk+\pp+\qq}u_\kk\left\{(R_1+J^{23}_{\kk,\pp,\qq})v_\pp \right. \nonumber \\
    &&\left.-[R_1\gamma_1(\pp+\qq)+J^{23}_{\kk,\pp,\qq}]u_\pp\right\} + u_{\kk+\pp+\qq}u_\kk R_1[\gamma_1(\kk)u_\pp  \nonumber \\
    && -\gamma_1(\kk+\pp+\qq)v_\pp] + v_{\kk+\pp+\qq}v_\kk R_1[\gamma_1(\kk+\pp+\qq)u_\pp  \nonumber \\
    &&-\gamma_1(\kk)v_\pp],  
    \end{eqnarray}
    \begin{eqnarray}
    &&\mathrm{f^{(\mathsf{d})}_{3m}}(\kk,\pp,\qq) = u_{\kk+\pp+\qq}v_\kk\left\{v_\pp(R_1+J^{23}_{\kk,\pp,\qq})-[R_1\gamma_1(\pp+\qq)\right.\nonumber \\
    &&\left.+J^{23}_{\kk,\pp,\qq}]u_\pp\right\} + v_{\kk+\pp+\qq}u_\kk\left\{(R_1\gamma_1(\pp+\qq)+J^{23}_{\kk,\pp,\qq})v_\pp \right. \nonumber \\
    &&\left.-(R_1+J^{23}_{\kk,\pp,\qq})u_\pp\right\} + u_{\kk+\pp+\qq}u_\kk R_1[\gamma_1(\kk+\pp+\qq)u_\pp  \nonumber\\
    &&-\gamma_1(\kk)v_\pp] + v_{\kk+\pp+\qq}v_\kk R_1[\gamma_1(\kk)u_\pp \nonumber\\
    &&-\gamma_1(\kk+\pp+\qq)v_\pp].
\end{eqnarray}
Here we define $J_{\kk,\pp,\qq}^{23}$ as
\begin{eqnarray}
J_{\kk,\pp,\qq}^{23}&&=z\sum\limits_{i=2,3}\left\{R_i[\gamma_i(\kk+\pp+\qq)+\gamma_i(\kk)\right. \notag\\
&&\left.-\gamma_i(\pp+\qq)-1]\right\}.
\end{eqnarray}
These equations are used in Eq.~\eqref{eq:i33m} for calculating Figs.~\ref{fig:fig11}(a) and \ref{fig:fig11}(b) in the main text.

Next, we will compute the second-order UCL contribution for the three magnon term. This term originates from the combination of a four-spin operator with a single spin operator. At $\mathcal{O}(\text{UCL}[2])$ order, given by Eq.~\eqref{eq:2ucl4s}, the RIXS operator matrix element is defined as
\begin{eqnarray}
    \mathrm{O}^{(2,S^{\frac{3}{2}},1)}_{\qq,\mathrm
{3m}}=\langle \mathrm
{3m}|\sum\limits_{i,j,l}e^{i\qq \cdot \br_i}\hat{S}_i^xh_i(\hat{\bs}_i\cdot \hat{\bs}_j)(\hat{\bs}_i\cdot \hat{\bs}_l)h_i^\dagger|g\rangle.\nonumber \\
\label{eq:o23m}
\end{eqnarray}
Similar to the process outlined for Eq.~\eqref{eq:4s}, there will be operators belonging to the $jl$ and $ij$ sites. The related RIXS operator matrix element can be written as
\begin{eqnarray}
\mathrm{O}^{(2_{jl},S^{\frac{3}{2}},1)}_{\qq,\mathrm{3m}}=	\langle \mathrm{3m}|\sum\limits_{i,j,l}e^{i\qq \cdot \br_i}\hat{S}^x_ih_i(\hat{\bs}_j \cdot \hat{\bs}_l)h_i^\dagger|g\rangle, \nonumber \\
\label{eq:o2jl3m}
\end{eqnarray}
whose Bogoliubov transformed version is
\begin{eqnarray}
			&&\mathrm{O}^{(2_{jl},S^{\frac{3}{2}},1)}_{\qq,\mathrm{3m}}=\langle \mathrm{3m}|
   \frac{S^\frac{3}{2}}{\sqrt{2N}}\sum\limits_{\kk,\pp}\left[\mathrm{f^{(\mathsf{e})}_{3m}}(\kk,\pp,\qq)\alpha^\dagger_\pp\alpha^\dagger_{\kk+\qq-\pp}\beta^\dagger_{-\kk} \right.\nonumber\\
 &&+\left.\mathrm{f^{(\mathsf{f})}_{3m}}(\kk,\pp,\qq)\beta^\dagger_{\pp}\alpha^\dagger_{\kk+\qq-\pp}\beta^\dagger_{-\kk}\right]|g\rangle+H.c..\nonumber\\
 \label{eq:o2p3mb}
\end{eqnarray} 
The three-magnon scattering matrix elements are given by 
\label{app:3msme}
\begin{flalign}
		&\mathrm{f^{(\mathsf{e})}_{3m}}(\kk,\pp,\qq)=f(\kk,\pp,\qq)(v_\pp u_{\kk+\qq-\pp}v_{\kk}-u_\pp v_{\kk+\qq-\pp}u_\kk),&\\
		&
        \mathrm{f^{(\mathsf{f})}_{3m}}(\kk,\pp,\qq)=f(\kk,\pp,\qq)(v_\pp v_{\kk+\qq-\pp}u_{\kk}-u_\pp u_{\kk+\qq-\pp}v_\kk),&
	\end{flalign}
with the following definition
\begin{eqnarray}
&&f(\kk,\pp,\qq)=-6[\cos(q_x-p_x)+\cos(q_y-p_y)]\notag\\
&&+2[\cos(2k_x+q_x-p_x)+\cos(2k_y+q_y-p_y)]\notag\\
&&\left. +4[\cos(k_x)\cos(k_y+q_y-p_y)+\cos(k_y)\cos(k_x+q_x-p_x)\right], \notag\\
\end{eqnarray}
which appears in Eq.~\eqref{eq:i43m}, Figs.~\ref{fig:fig6}(c) and \ref{fig:fig6}(d). The $ij$ combination RIXS operator at this order is \begin{eqnarray}
\mathrm{O}^{(2_{ij},S^{\frac{3}{2}},1)}_{\qq,\mathrm{3m}}=	\langle \mathrm{3m}|\sum\limits_{i,j}e^{i\qq \cdot \br_i}\hat{S}^x_ih_i(\hat{\bs}_i \cdot \hat{\bs}_j)h_i^\dagger|g\rangle.\nonumber \\
\label{eq:o2ij3m}
\end{eqnarray}

\subsection{SBMFT RIXS operator and scattering matrix elements}

\label{app:ssme}

In this section, we document the explicit expressions of the SBMFT RIXS mean-field spinon scattering operators and scattering matrix elements. Considering spinon BEC the SBMFT RIXS scattering matrix elements of each mean-field spinon will have two parts: the continuum part (denoted with $\mathrm{c}$) and the singular part (denoted with $\mathrm{s}$). The singular part arises from BEC spinon condensation at $\pm \frac{\bq}{2}$. The scattering operators that span over the other wave vectors belong to the continuum part. The Bogoliubov version of the SBMFT two spin-flip RIXS operator matrix element [Eq.~\eqref{eq:o2smf}] can be written for the mean-field two-spinon (denoted by $\mathrm{2s}$) RIXS operator matrix element as
\begin{eqnarray} &&\mathrm{O}^{SB}_{\qq,\mathrm{2s}}=\nonumber\\
&&\langle \mathrm{2s}|\sum\limits_{\kk}\mathrm{f_{2sc}}(\kk,\qq)(\alpha^{s\dagger}_{\kk+\frac{\qq}{2}}\beta^{s\dagger}_{-\kk+\frac{\qq}{2}}-\beta^{s\dagger}_{\kk+\frac{\qq}{2}}\alpha^{s\dagger}_{-\kk+\frac{\qq}{2}})|\mathcal{GS}\rangle_{\mathrm{SB}}\nonumber\\
&&+H.c..
\label{eq:o2sb}
   \end{eqnarray}
The scattering matrix element $\mathrm{f_{2sc}}(\kk,\qq)$ of the mean-field two-spinon continuum part is     
					\begin{eqnarray}
			&\mathrm{f_{2sc}(\kk,\qq)}=\frac{z}{2}A_{\delta}(\gamma^A_{\kk+\frac{\qq}{2}}+\gamma^A_{\kk-\frac{\qq}{2}})(u^s_{\kk+\frac{\qq}{2}}u^s_{\kk-\frac{\qq}{2}}+v^s_{\kk+\frac{\qq}{2}}v^s_{\kk-\frac{\qq}{2}}).\nonumber\\
		\end{eqnarray}
Considering spinon BEC which leads to $u_{\pm\frac{\bq}{2}}\sim |v_{\pm\frac{\bq}{2}}|\sim \sqrt{\frac{N m_0}{2}}$, $\omega^s_{\pm\frac{\bq}{2}}\sim 0$~\cite{PhysRevB.91.134423,auerbach2012interacting}, the scattering matrix elements of the singular part are 
\begin{eqnarray}
&&\mathrm{f_{2ssp}(\qq)}=\frac{z}{2}A_{\delta}(\gamma^A_{\frac{\bq}{2}-\frac{\qq}{2}}+\gamma^A_{\frac{\bq}{2}+\frac{\qq}{2}})(u^s_{\frac{\bq}{2}+\frac{\qq}{2}}u^s_{\frac{\bq}{2}-\frac{\qq}{2}} \nonumber \\
&&+v^s_{\frac{\bq}{2}+\frac{\qq}{2}}v^s_{\frac{\bq}{2}-\frac{\qq}{2}}), \\
&&\mathrm{f_{2ssm}(\qq)}=\frac{z}{2}A_{\delta}(\gamma^A_{-\frac{\bq}{2}-\frac{\qq}{2}}+\gamma^A_{-\frac{\bq}{2}+\frac{\qq}{2}})(u^s_{-\frac{\bq}{2}+\frac{\qq}{2}}u^s_{-\frac{\bq}{2}-\frac{\qq}{2}} \nonumber\\
&&+v^s_{-\frac{\bq}{2}+\frac{\qq}{2}}v^s_{-\frac{\bq}{2}-\frac{\qq}{2}}),
	\end{eqnarray}
where $\mathrm{2ssp}$ and $\mathrm{2ssm}$ are the mean-field two-spinon singular part $\pm\frac{\bq}{2}$. These matrix elements are used in Eqs.~\eqref{eq:ic2s}, \eqref{eq:is2s}, and Figs.~\ref{fig:fig8}(a) and \ref{fig:fig8}(b). 
 
The Bogoliubov transformed Eq.~\eqref{eq:o4s} of the mean-field four-spinon (denoted by $\mathrm{4s}$) RIXS operator matrix element is given by
\begin{eqnarray}
 \notag   &&\mathrm{O}^{SB}_{\qq,\mathrm{4s}}=\langle \mathrm{4s}|\sum\limits_{\kk,\pp,\pp'}\mathrm{f_{4sc}}(\kk,\pp,\pp',\qq)(\alpha_\pp^{s\dagger}\beta^{s\dagger}_{\pp'}+\beta_\pp^{s\dagger}\alpha^{s\dagger}_{\pp'})\\
    &&(\alpha^{s\dagger}_{\kk+\qq}\beta^{s\dagger}_{-\kk-\pp-\pp'}-\beta^{s\dagger}_{\kk+\qq}\alpha^{s\dagger}_{-\kk-\pp-\pp'}) |\mathcal{GS}\rangle_{\mathrm{SB}}+H.c.. \nonumber\\
    \label{eq:o4sb}
\end{eqnarray}
The scattering matrix elements of the continuum part of the four-spinon (denoted by $\mathrm{4sc}$) case is
\begin{eqnarray}
	&&\mathrm{f_{4sc}(\kk,\pp,\pp',\qq)}=A_{\delta}(\gamma^A_{\kk+\pp+\pp'}+\gamma^A_{\kk+\qq})
    (u^s_\pp v^s_{\pp'}+u^s_{\pp'}v^s_\pp) \nonumber\\
    &&(u^s_{\kk+\qq}u^s_{\kk+\pp+\pp'}+v^s_{\kk+\qq}v^s_{\kk+\pp+\pp'}),
		\end{eqnarray}
while the singular part is
\begin{flalign}
	&\mathrm{f_{4ssp}(\qq)}=A_{\delta}(\gamma^A_{\frac{3\bq}{2}}+\gamma^A_{\frac{\bq}{2}+\qq})(u^s_{\frac{\bq}{2}+\qq}+v^s_{\frac{\bq}{2}+\qq}),\\
 &\mathrm{f_{4ssm}(\qq)}=A_{\delta}(\gamma^A_{\frac{-3\bq}{2}}+\gamma^A_{-\frac{\bq}{2}+\qq})(u^s_{-\frac{\bq}{2}+\qq}+v^s_{-\frac{\bq}{2}+\qq}).
		\end{flalign}
Here, $\mathrm{4ssp}$ and $\mathrm{4ssm}$ are the mean-field four-spinon singular part at $\pm\frac{\bq}{2}$. These two equations are used in Eqs.~\eqref{eq:i4sc} and ~\eqref{eq:i4ss} in the main text. 

\section{RIXS intensity}\label{app:ri}
We can derive the zero-temperature Green's function from RIXS operator matrix elements. For example, the two-magnon propagator can be obtained from $\mathrm{O}^{(1,S,0)\dagger}_{\qq,\mathrm{2m}}(t)\mathrm{O}^{(1,S,0)}_{\qq,\mathrm{2m}}(0)$ as 
\begin{align}
     &G^{(1)}_{\mathrm{2m}}(\qq,t)=-i \left[S \mathrm{f^{(1)}_{2m}}(\kk,\qq)\right]^2 \nonumber\\
 &\hat{T}\langle g|\alpha_{\kk+\frac{\qq}{2}}(t)\alpha^\dagger_{\kk+\frac{\qq}{2}}(0)\beta_{-\kk+\frac{\qq}{2}}(t)\beta^\dagger_{-k+\frac{\qq}{2}}(0)|g\rangle,
\end{align} where the matrix element $\mathrm{O}_\qq(t)$ depends on the momentum $\qq$ and time $t$. The time ordering operator is given by $\hat
     {T}$.  After applying Wick's theorem we can identify the two-magnon RIXS intensity as
\begin{align}
&I^{(1,S^2,0)}_{\qq,\mathrm{2m}}(\qq,\omega)=-\frac{1}{\pi}\mathrm{Im}[G^{(1)}_{\mathrm{2m}}(\qq,\omega)]\nonumber\\
&=
\frac{1}{N}\sum\limits_{\kk}\left[S \mathrm{f^{(1)}_{2m}}(\kk,\qq)\right]^2\delta(\omega-\omega_{\kk+\frac{\qq}{2}}-\omega_{\kk-\frac{\qq}{2}}),
\end{align}
which is Eq.~\eqref{eq:i12m} in the main text. For the other types of two-magnon RIXS operator matrix element Eq.~\eqref{eq:o2p2m}, we can write the Green's function at $\mathcal{O}(\text{UCL}[2])$ as
  \begin{align}
     &G^{(2_{jl})}_{\mathrm{2m}}(\qq,t)=-i \left[S \mathrm{f^{(2_{jl})}_{2m}}(\kk,\qq)\right]^2 \nonumber\\
 &\hat{T}\langle g|\alpha_{\kk}(t)\alpha^\dagger_{\kk}(0)\beta_{-\kk+\qq}(t)\beta^\dagger_{-k+\qq}(0)|g\rangle.
\end{align} 
The above can be used to derive Eq.~\eqref{eq:i22m}. We will have contributions from the cross terms between Eq.~\eqref{eq:o1s} and Eq.~\eqref{eq:o2p2m} to obtain 
\begin{align}
     &G^{(2_{ij},2_{jl})}_{\mathrm{2m}}(\qq,t)= \nonumber\\
&-i\left[S^2 \mathrm{f^{(2_{ij})}_{2m}}(\kk,\qq)\mathrm{f^{(2_{jl})}_{2m}}(\kk,\qq)\right]\nonumber\\
&\hat{T}\left[\langle g|\alpha_{\kk+\frac{\qq}{2}}(t)\alpha^\dagger_{\kk}(0)\beta_{-\kk+\frac{\qq}{2}}(t)\beta^\dagger_{-\kk+\qq}(0) \right.\nonumber\\
&\left.+\alpha_{\kk}(t)\alpha^\dagger_{\kk+\frac{\qq}{2}}(0)\beta_{-\kk+\qq}(t)\beta^\dagger_{-\kk+\frac{\qq}{2}}(0)|g\rangle\right].
\label{eq:ct}
\end{align}
Following the procedure outlined above, the cross term in the two-magnon RIXS intensity expression is given by 
 \begin{align}
&I^{([2_{ij},2_{jl}],S^2,0)}_{\qq,\mathrm{2m}}(\qq,\omega)=-\frac{1}{\pi}\mathrm{Im}[G^{(2_{ij},2_{jl})}_{\mathrm{2m}}(\qq,\omega)]\nonumber\\
&=\frac{S^2}{N}\sum\limits_{\kk}\left[ \mathrm{f^{(2_{ij})}_{2m}}(\kk,\qq)\mathrm{f^{(2_{jl})}_{2m}}(\kk,\qq)\right]\nonumber\\
&[\delta(\omega-\omega_{\kk+\frac{\qq}{2}}-\omega_{\kk-\frac{\qq}{2}})+\delta(\omega-\omega_{\kk}-\omega_{\kk-\qq})],
\end{align}
which is Eq.~\eqref{eq:i122m}. The procedures outlined above can be applied to recover the remainder of the RIXS intensity expressions stated in the text and extended to the three-magnon cases.

\onecolumngrid
\section{supplemental material}\label{app:supp}
The acronyms, notations, and mathematical symbols used in the main article are summarized in the following two tables.

\begin{center}
\begin{table}[H]
\centering
\caption{Explanation of acronyms and notations.}
\label{tab:tab1}
\resizebox{0.65\linewidth}{!}{
\begin{tabular}{|c|l|}
\hline
\textbf{Acronyms}/\textbf{Notations} & \textbf{Significance} \\
\hline
LSWT & linear spin wave theory  \\
HP & Holstein-Primakoff   \\
SBMFT & Schwinger boson mean-field theory \\
UCL & ultrashort core-hole lifetime  \\
SB & Schwinger boson \\
MF &  mean-field \\
(N)SC & (non-)spin-conserving  \\
RVB & resonating valence bond  \\
MBZ & magnetic Brillouin Zone  \\
DOS & density of states  \\
RIXS & resonant inelastic x-ray scattering  \\
INS & inelastic neutron scattering  \\
DSF & dynamical structure factor   \\
CUT & continuous unitary transformation \\
CST & continuous similarity transformation  \\
VBC & valence bond crystal    \\
AF & antiferromagnet \\
 SOC & spin-orbit coupling \\
 BEC & Bose-Einstein condensate \\
 IGG &  Invariant Gauge Group \\
  mm & multimagnon  \\
 tot &  total \\
 $\mathrm{1m}$ & single-magnon  \\
 $\mathrm{2m}$ &  two-magnon \\
$\mathrm{bm}$  & bimagnon  \\
 $\mathrm{3m}$ &  three-magnon  \\
 $\mathrm{2s}$  & mean-field two-spinon\\
 $\mathrm{4s}$ &  mean-field four-spinon\\
\hline
\end{tabular}
}
\end{table}
\end{center}

\onecolumngrid
\begin{center}
\begin{table}[H]
\label{tab:tab2}
\caption{Explanation of mathematical symbols.\\
The acronyms used in this table are defined in Table~\ref{tab:tab1}.}
\resizebox{\linewidth}{!}{
\begin{tabular}{|c|l|c|l|}
\hline
\textbf{Symbols} & \textbf{Significance} &
\textbf{Symbols} & \textbf{Significance} \\
\hline
$\bq$ & ordering wave vector &
$\kk_{in(out)}$ & the incident (outgoing) photon momentum \\
$\epsilon(\epsilon^{\prime})$ & the incident (outgoing) photon polarization  &
$|\kk_{in},\epsilon\rangle$ & the initial state in the RIXS process \\
$|\kk_{out},\epsilon^{\prime}\rangle$ & the final state in the RIXS process &
$\qq$ & scattering wave vector \\
$|g\rangle$ & the quantum-corrected N\'{e}el state &
$|c\rangle$ & the averaged RVB state \\
$|m\rangle$ & the mm eigenstate &
$|s\rangle$ & the mean-field spinon eigenstate \\
$|2s\rangle$ & the 2s excited state &
$|4s\rangle$ & the 4s excited state \\
$|\mathcal{GS}\rangle_{\textrm{swt}}$ & spin wave theory ground state&
$|\mathcal{GS}\rangle_{\mathrm{SB}}$ & the SB ground state \\
$N$ & the number of lattice sites &
$m_0$ & magnetization in SBMFT \\
$\kappa$ & spin-wave dispersion coefficient &
$\epsilon$ & spin-wave dispersion coefficient\\
$\omega$ & energy (dispersion) of spin-wave &
$\omega^s$ & energy (dispersion) of mean-field spinon\\
 $\gamma$ & structure factor  &
$a(b)$ &  HP boson operator\\
$a^s(b^s)$ &  spinon operator in terms of SB&
$\phi$ & spinor\\
$|0\rangle_{\mathrm{SB}}$ &  SB vacuum &
$|\uparrow \rangle $& spin-up $S=\frac{1}{2}$ state \\
$|\downarrow \rangle $ & spin-up $S=\frac{1}{2}$ state  &
$\hat{A}_{ij}$ & singlet bond operator \\
$\langle \hat{A}_{ij}\rangle$ & mean-field value of singlet bond operator &
$A_{\delta}$ & mean-field order parameter \\
$\lambda$ & local constraint field &
$\alpha(\beta)$ & Bogoliubov quasiparticles of HP bosons \\
$\alpha^s(\beta^s)$ & Bogoliubov quasiparticles of SB &
$u(v)$ & Bogoliubov coefficients of HP bosons \\
$u^s(v^s)$ & Bogoliubov coefficients of SB  & $\mathcal{L}$ & the order of UCL expansion \\
$\mathcal{M}$ & the spin-wave order of  RIXS scattering operators   &
$\mathcal{N}$ & the order of single spin-flip operator  \\
$\xi$ & $\tilde{J}_1/\Gamma$, the leading term of the UCL coefficient &
$\Gamma$ & the inverse core-hole lifetime \\
$\mathcal{O}(\text{UCL}[\mathcal{L}])$ & $\mathcal{L}$-th order of the UCL expansion &
$W_{nm}^{{(\mathcal{L},\mathcal{M}^2,\mathcal{N})}}$ & RIXS spectral weight of mm \\
$\mathrm{O}^{(\mathcal{L},\mathcal{M},\mathcal{N})}_{\qq}$  & general form of RIXS operator matrix element & $\mathrm{O}^{SB}_{\qq,\mathrm{2s}}$  & 2s RIXS operator matrix element \\
$\mathrm{O}^{SB}_{\qq,\mathrm{4s}}$  & 4s RIXS operator matrix element & $\mathrm{O}^{(0,\sqrt{S},1)}_{\qq,\mathrm{1m}}$  & 1m RIXS operator matrix element \\
$\mathrm{O}^{(\mathcal{L},S,0)}_{\qq,\mathrm{2m}}$  & 2m RIXS operator matrix element in $\mathcal{O}(\text{UCL}[\mathcal{L}])$& $\mathrm{O}^{(\mathcal{L},S^{\frac{3}{2}},1)}_{\qq,\mathrm{3m}}$  & 3m RIXS operator matrix element in $\mathcal{O}(\text{UCL}[\mathcal{L}])$ \\
$\mathrm{f^{(p)}_{2m}}$ & 2m scattering matrix element, $p=1,2_{jl},[2_{ij},2_{jl}]$  & $\mathrm{f^{(p)}_{3m}}$ & 3m scattering matrix element, $p=\mathsf{a},\mathsf{b},\mathsf{c},\mathsf{d},\mathsf{e},\mathsf{f}$ \\
$\mathrm{f_{2sc}}$ & scattering matrix element of the continuum part of 2s  &  
$\mathrm{f_{4sc}}$ & scattering matrix element of the continuum part of 4s   \\
$\mathrm{f_{2ssp}}$ & scattering matrix element of the singular part of 2s at $\frac{\bq}{2}$& $\mathrm{f_{4ssp}}$ & scattering matrix element of the singular part of 4s at $\frac{\bq}{2}$\\
$\mathrm{f_{2ssm}}$ & scattering matrix element of the singular part of 2s at $-\frac{\bq}{2}$& $\mathrm{f_{4ssm}}$ & scattering matrix element of the singular part of 4s at $-\frac{\bq}{2}$\\
$I^{(\mathcal{L},\mathcal{M}^2,\mathcal{N})}$  & general form of RIXS intensity&
$\mathcal{I}_{mm}$  & mm RIXS intensity\\
$I^{(0,S,1)}_{\mathrm{1m}}$  & 1m RIXS intensity&
$I^{(\mathcal{L},S^2,1)}_{\mathrm{2m}}$ & 2m RIXS intensity in $\mathcal{O}(\text{UCL}[\mathcal{L}])$\\
$I^{(\mathcal{L},S^3,1)}_{\mathrm{3m}}$ & 3m RIXS intensity in $\mathcal{O}(\text{UCL}[\mathcal{L}])$&
$I_{3m}^{{(0,1/S,1)}}$ & RIXS intensity of one-to-three magnon hybridization \\
$I^{\text{tot}}_{\mathrm{2m}}$ & 2m RIXS intensity up to $\mathcal{O}(\text{UCL}[2])$&
$I^{S^3,\text{tot}}_{\mathrm{3m}}$ & 3m RIXS intensity up to $\mathcal{O}(\text{UCL}[2])$\\
$I^{c}_{\mathrm{2s}}$ & continuum part of 2s RIXS intensity &
$I^{s}_{\mathrm{2s}}$ & singular part of 2s RIXS intensity\\
$I^{c}_{\mathrm{4s}}$ & continuum part of 4s RIXS intensity &
$I^{s}_{\mathrm{4s}}$ & singular part of 4s RIXS intensity\\
$D_{\mathrm{2m}}$  & DOS of 2m&
$D_{\mathrm{3m}}$ & DOS of 3m\\
$D_{\mathrm{4s}}$ & DOS of 4s&
$\hat{H}_{MF}$ & SBMFT Hamiltonian with mean-field ansatz\\
H & total system Hamiltonian &
$\mathcal{H}_0$ & spin-wave Hamiltonian of ground state \\
$\mathcal{H}_1$ & spin-wave Hamiltonian in $1/S$ order &
$V^{(p)}_{1234}$& vertex functions, $p =1,2,3,4,5,6$  \\
$\mathcal{V}$ & vertex function of one-to-three magnon hybridization &
$\hat{v}$ & bm scattering channel \\
$\hat{\Gamma}$ & bm interacting potentials &
$G(\kk,\omega)$ & momentum space Green's function \\
$G^{(\mathcal{L})}_{\mathrm{2m}}(\kk,\omega)$ & momentum space 2m Green's function in $\mathcal{O}(\text{UCL}[\mathcal{L}])$  &
 $F$ & the free energy\\
$\hat{g}$ & bm channel Green's function&
$\hat{R}$ & bm inter-channel propagator \\
$\Pi^{(1)}_{2m}$, $\Pi^{(2_{jl})}_{2m}$ & non-interacting 2m propagator &
$c_1,c_2,c_3$ & coefficients of 2$^{\text{nd}}$ order UCL expansion \\
\hline
\end{tabular}
}
\end{table}
\end{center}

\twocolumngrid
\bibliography{ref}
\end{document}